\documentclass[12pt,a4paper]{iopart}
\usepackage{iopams}  
\usepackage{graphicx,psfrag,bbm,latexsym,color,dcolumn,bm,dsfont,bbm}

\newcommand{\ua}{\uparrow}
\newcommand{\da}{\downarrow}
\newcommand{\media}[1]{\langle #1 \rangle}
\newcommand{\E}{\mbox{$\mathsf E$}}

\begin{document}

\title[Exact Monte Carlo time dynamics in many-body lattice quantum systems]
{Exact Monte Carlo time dynamics in many-body lattice quantum systems}

\author{Massimo Ostilli$^{1,2}$ and Carlo Presilla$^{1,2,3}$}

\address{$^1$\ Dipartimento di Fisica, Universit\`a di Roma ``La Sapienza'',
Piazzale A. Moro 2, Roma 00185, Italy}
\address{$^2$\ Center for Statistical Mechanics and Complexity, Istituto Nazionale per la Fisica della Materia, Unit\`a di Roma 1, Roma 00185, Italy}
\address{$^3$\ Istituto Nazionale di Fisica Nucleare, Sezione di Roma 1, 
Roma 00185, Italy}

\date{\today}

\begin{abstract}
On the base of a Feynman-Kac--type formula involving Poisson stochastic 
processes, recently a Monte Carlo algorithm has been introduced,
which describes exactly the real- or imaginary-time evolution of 
many-body lattice quantum systems. 
We extend this algorithm to the exact simulation of time-dependent correlation 
functions. 
The techniques generally employed in Monte Carlo simulations 
to control fluctuations, namely reconfigurations and importance sampling,
are adapted to the present algorithm and their validity is rigorously proved.
We complete the analysis by several examples for the hard-core boson 
Hubbard model and for the Heisenberg model.
\end{abstract}

\pacs{02.50.-r, 02.70.Ss}
\maketitle

\section{Introduction}
Probabilistic techniques, as the Quantum Monte Carlo (QMC) algorithms, 
provide an essential tool to investigate the properties of many-body systems.
Basically, with these techniques one evaluates functions of matrices 
by a random walk in the space of the matrix indices \cite{LINDEN}.
Given the Hamiltonian $H$ of the system, the idea is to find 
an appropriate stochastic representation
of the imaginary time evolution operator $U(t)=\exp(-Ht)$ applied to some 
initial trial state.
By using these methods, one can obtain, 
at least in the absence of a sign problem, 
the ground-state properties with a numerical effort that grows 
with some power of the size $L$ of the system.
On the other hand, the exact diagonalization of the Hamiltonian 
would imply an effort exponentially growing with $L$.

From a more physical point of view, a probabilistic representation 
provides a dual picture of the quantum systems:
on one hand, the traditional description in terms of bra, ket and operators,
on the other hand, a description in terms of expectations of suitable 
stochastic functionals,
which are averages over virtual trajectories of the particles.
It is this mapping with a (in a sense) classical system that allows us 
to extract quantum information by statistical simulations. 

In recent years, it has been proved that the dynamics of a system of
quantum particles on a lattice admits an exact probabilistic representation
\cite{DAJLS,DJS,PRESILLA}. 
A suitable stochastic functional $\mathcal{M}^{[0,t)}_{\bm{n}_0}$, 
which is defined in terms of a collection of independent Poisson processes 
and diffuses from a Fock state $\bm{n}_0$ to a Fock state $\bm{n}_t$, 
has the property that the expectation value 
$\E (\mathcal{M}^{[0,t)}_{\bm{n}_0}\delta_{\bm{n}_t,\bm{n}})$,
taken with respect to the Poisson processes, 
gives the matrix element of $U(t)$ between the two Fock states  
$\bm{n}_0$ and $\bm{n}$.
In the theory of stochastic processes, this probabilistic representation 
may be regarded as the lattice version of the Feynman-Kac formula. 
We emphasize that in this method no approximation
is introduced and no ``infinity path integral extrapolation'' is requested. 
It will be referred to as the exact probabilistic representation
(EPR) of the evolution operator $U(t)$.
The validity of EPR is not limited to Hamiltonian systems: 
it can be used to express any imaginary- or real-time evolution operator 
$U(t)$ having any finite matrix $H$ as generator.

In Ref.~\cite{PROST} we used EPR to obtain semi-analytical results 
in the limit $t\to \infty$, in which a central limit theorem applies. 
In this paper, we consider EPR at arbitrary times 
within a Monte Carlo approach (EPRMC).

Two other well known QMC algorithms, 
namely the path integrals Monte Carlo method (PIMC) and 
the Green function Monte Carlo method (GFMC),
have affinities with EPRMC and a comparison is mandatory.

In PIMC, one evaluates the matrix elements of $U(t)$
by using the Trotter approximation \cite{CEPERLEYKALOS}.
The operator $U(t)$ is factorized in the kinetic, 
$\exp(-Tt)$, and interaction, $\exp(-Vt)$, terms
so that one gets
$\exp(-Ht)=\prod_{k=1}^{N}\exp(-Tt/N)\exp(-Vt/N)+\mathcal{O}([T,V]t^2/N^{2})$.
This approximation leads to a Feynman-Kac formula, 
in which, as in EPRMC, the trajectories in the Fock space are generated 
only by the kinetic part, $\exp(-Tt/N)$.
However, in contrast to EPRMC, 
there are no stochastic times related to Poisson processes.
The PIMC simulations are performed by evolving 
the trajectories for $N$ steps. 
The drawback is that to obtain results corresponding to $t/N \to 0$, 
in which the Trotter approximation becomes exact, 
one must use extrapolation procedures. 
For any finite value of $N$, the extrapolation becomes unreliable 
for values of $t$ sufficiently large.
This is particularly evident in the case of real times ($t \rightarrow it$), 
when the matrix elements of $U$ have an oscillating behavior 
with respect to $t$.
In contrast, no small step approximation is introduced in EPRMC
and no extrapolation is requested.

Now, let us consider GFMC. 
The method consists in repeated statistical applications of the operator 
$(1-Ht/N)$ to an initial state.
For $N\to \infty$, one reproduces the operator $U(t)$,
whereas an approximation affected by a relative error $\varepsilon (N)$
is obtained for any finite $N$. 
It is plausible that the sampling directly the operator $U(t)$ 
instead of $(1-Ht/N)^N$ leads to a higher efficiency.
In the Appendix, we show that the relative efficiency
between EPRMC and GFMC in filtering the ground state
is $E_0^{2}/[2E_{0}^{(0)}(E_1-E_0)\varepsilon]$, 
where $E_0$ and $E_1$ are the energies of the ground- and first-excited 
states of the considered system and $E_{0}^{(0)}$ is 
the ground-state energy of the associated non interacting system.
Since the gap $(E_1-E_0)$ decreases as the size $L$ of the system 
is increased, compared to GFMC, EPRMC offers a more powerful method 
to investigate the ground-state properties of large lattice systems.

Actually, the GFMC scheme mentioned above is rather crude.
Trivedi and Ceperley \cite{TRIVEDICEPERLEY} introduced Poisson processes 
as a tool to obtain a more efficient GFMC method when the transition
probabilities, proportional to the matrix elements 
of $Ht/N$, vanish for $N \to \infty$. 
We will refer to this improved GFMC as GFMCP.
In Ref.~\cite{PRESILLA} it has been proved that in the limit $N \to \infty$ 
GFMCP becomes equivalent to EPRMC.
However, as explained in the Appendix, for a finite value of $N$
the relative efficiency of EPRMC with respect to GFMCP is 
$(E_0/E_0^{(0)})^2/ 2\varepsilon$, \textit{i.e.} it is proportional
to the accuracy $\varepsilon^{-1}$ required in the approximated GFMCP.

Controlling the large fluctuations is one of the most important issues
of any Monte Carlo method.
This is evident in GFMC where an iterated statistical 
application of the operator $(1-Ht/N)$ is performed.
Roughly speaking, after $k$ iterations one has 
fluctuations that grow like $\Delta^k$, 
$\Delta$ being the statistical error associated to a single step. 
To solve the problem of large fluctuations, 
besides the development of the importance sampling method \cite{CEPERLEYKALOS},
remarkable progress has been made with the reconfiguration
technique introduced by Hetherington \cite{HETHERINGTON} 
and subsequently improved by Sorella \cite{CALANDRASORELLA}
who proposed a scheme without bias (see also Ref.~\cite{BECCARIA}).

In this paper, after introducing some relevant physical models 
(Section \ref{Models})
and recalling EPR (Section \ref{Probabilistic representation}), 
we extend EPR to the study of exact time-dependent correlation functions 
(Section \ref{Correlation functions}). 
In the core Section \ref{EPRMC algorithm}, we discuss the 
EPRMC algorithm first with a pure sampling and then adding 
fluctuation control by reconfigurations and importance sampling.
A detailed proof of the validity of the reconfiguration method 
is given in Section \ref{Proof of the reconfiguration method}.
Results of numerical simulations for the hard-core boson Hubbard model 
and for the Heisenberg model are discussed in Section \ref{Numerical results}.
Conclusions are drawn in Section \ref{Conclusions}.

\section{Models}
\label{Models}
The Hamiltonian models of interest have the following general structure
(we shall always take $\hbar=1$) 
\begin{eqnarray}
\label{Hamiltoniana}
{H} = {T}+{V}, 
\end{eqnarray}
where $
{V}$ is the potential energy operator and ${T}$ the kinetic energy operator, 
which on a lattice assumes the form
\begin{eqnarray}
\label{Hubbard}
{T} &=& - 
\sum_{i < j \in \Lambda} 
\sum_{\sigma=\ua\da} \eta_{ij}
\left( 
c^\dag_{i\sigma} c^{}_{j\sigma} + c^\dag_{j\sigma} c^{}_{i\sigma}
\right) .
\end{eqnarray}
Here $\Lambda\subset Z^d$ is a finite $d$-dimensional lattice
with $|\Lambda|$ ordered sites 
and $c_{i \sigma}$ the commuting or anticommuting 
destruction operators at site $i$ and spin index $\sigma$
with the property $c_{i\sigma}^2=0$ (fermion or hard-core boson systems).
The system is described in terms of Fock states labeled by 
the configuration
$\bm{n}= (n_{1 \ua},n_{1 \da}, \ldots, n_{|\Lambda| \ua}, n_{|\Lambda| \da})$, 
\textit{i.e.} the set of lattice occupation numbers taking the values 0 or 1. 
The total number of particles is
$N_\sigma = \sum_{i \in \Lambda} n_{i\sigma}$ for $\sigma=\ua \da$.
We shall use the mod 2 addition $n \oplus n'= (n+n') \bmod 2$.

The analysis we develop in the following 
is valid for an arbitrary functional form of the potential ${V}$. 
However, numerical examples will be limited to the well known Hubbard 
potential \cite{HUBBARD} 
\begin{eqnarray}
V=\sum_{i \in \Lambda} \gamma_i 
c^\dag_{i\ua} c^{}_{i\ua}~c^\dag_{i\da} c^{}_{i\da}.
\end{eqnarray}
We emphasize that, independently of its form, $V$ is diagonal 
in the Fock space, whereas ${T}$ is off diagonal.

In this paper we will consider only hard-core boson systems.
We recall that, even if fermion systems, 
like the Hubbard model, are more attractive, hard-core bosons have not a 
purely academic interest.
Besides the description of boson particles with a hard-core interaction, they can be mapped onto
systems of half integer spin \cite{LINDEN,TRIVEDICEPERLEY,MATSUBARA}.
As an example, we consider the $S=\frac{1}{2}$ Heisenberg quantum 
antiferromagnetic model  
\begin{eqnarray}
\label{HEISENBERG}
H=J\sum_{\media{i,j}}\bm{S}_{i}\cdot\bm{S}_{j}=\frac{J}{2} 
\sum_{\media{i,j}}(S^{+}_{i}S^{-}_{j}+S^{-}_{i}S^{+}_{j})+
J\sum_{\media{i,j}}S^{z}_{i}S^{z}_{j},
\end{eqnarray}
where $J>0$, 
$\media{i,j}$ indicates that the sites $i$ and $j$ are nearest neighbors, 
and $\bm{S}_{i}$ and $\bm{S}_{j}$ are the spin operators. 
It is convenient to view the left and right factors in 
$\bm{S}_{i}\cdot\bm{S}_{j}$ as the spin operators of two sublattices 
A and B, respectively. 
The mapping is then established as follows. 
The operators $S_{i}^{+}$ and $S_{j}^{-}$ commute on different sites
and are thus identified with boson operators via 
$b_{i}^{\dag}=S_{i}^{+}$ and $b_{j}=S_{j}^{-}$.
Furthermore as 
$S_{i}^{z}=S_{i}^{+}S_{i}^{-}-\frac{1}{2}$, 
one has $S_{i}^{z}=n_{i}-\frac{1}{2}$, where $n_{i}=b_{i}^{\dag}b_{i}$ 
is the number operator. 
For a half spin system
$S_{i}^{+}S_{i}^{+}=S_{i}^{-}S_{i}^{-}=0$, 
which implies $(b^{\dag}_{i})^{2}=0$. 
Therefore, the bosons have a hard core and
a site can be occupied by at most one particle. 
In order to a have negative sign in the kinetic energy term, 
a further transformation is necessary. 
The spins on the sublattice B are rotated as 
$S_{j}^{x}\rightarrow -S_{j}^{x}$,
$S_{j}^{y}\rightarrow -S_{j}^{y}$, and $S_{j}^{z}\rightarrow S_{j}^{z}$. 
Since this transformation is unitary,
the commutation relations are left unchanged.
The hard-core boson Hamiltonian then reads
\begin{eqnarray}
\label{HEISENBERG1}
H=-\frac{J}{2}\sum_{\media{i,j}}(b_{i}^{\dag}b_{j}+b_{j}^{\dag}b_{i})+
J\sum_{\media{i,j}}n_{i}n_{j}+E_{N},
\end{eqnarray}
where $E_{N}={-JZ|\Lambda|}/8$, $Z$ being the number of nearest neighbors
for the given lattice, 
\textit{e.g.} $Z=4$ for a square lattice in two dimensions.

\section{Probabilistic representation}
\label{Probabilistic representation}
We are interested in evaluating matrix elements of the form
$\langle \bm{n} |e^{- Ht} | \bm{n}_0 \rangle$ or
$\langle \bm{n} |e^{- i Ht} | \bm{n}_0 \rangle$
between two Fock states $\bm{n}_0$ and $\bm{n}$ with $t\in \mathbb{R}$.
As usual, we will speak of imaginary times in the former case and
of real times in the latter one. 

Let $\Gamma$ be the set of links, \textit{i.e.}
the pairs $(i,j)$ with $1 \leq i<j \leq |\Lambda|$
such that  $\eta_{ij}\neq 0$. 
For simplicity, we will assume $\eta_{ij}=\eta$  
if $i$ and $j$ are first neighbors and $\eta_{ij}=0$ otherwise.
%We will also assume $\gamma_i=\gamma$.
For a $d$-dimensional lattice the number of links per spin component is 
$|\Gamma| = d |\Lambda|$.
Let us introduce
\begin{eqnarray}
\label{lambda}
\lambda_{ij \sigma}(\bm{n}) &\equiv& 
\langle \bm{n} \oplus \bm{1}_{i\sigma} \oplus \bm{1}_{j\sigma}|
c^\dag_{i\sigma} c^{}_{j\sigma} + c^\dag_{j\sigma} c^{}_{i\sigma}
|\bm{n}\rangle ,
\\
\label{potential}
V(\bm{n}) &\equiv& 
\langle \bm{n}|H|\bm{n}\rangle ,
\end{eqnarray}
where $\bm{1}_{i\sigma}=(0,\ldots,0,1_{i\sigma},0,\ldots,0)$.
Note that the values assumed by $\lambda_{ij \sigma}$
are 0 or 1 ($\lambda_{ij \sigma}=0,\pm1$ is possible in the case of fermion 
systems not considered here).
We will call the link $(ij\sigma)$ active if $\lambda_{ij\sigma} \neq 0$.
Let $\{N^{t}_{ij\sigma}\}$, $(i,j) \in \Gamma$, be a family
of $2|\Gamma|$ independent left continuous Poisson processes with 
jump rate $\rho=\eta$
if $\lambda_{ij\sigma} \neq 0$ and $0$ otherwise \cite{NOTE}.
Let us now define the stochastic dynamics on the lattice.
At each jump of the process $N_{ij\sigma}^{t}$ 
a particle with spin $\sigma$ moves from site $i$ to site $j$
or \textit{vice versa}. 
Let us indicate with $A(\bm{n})$ the number of active links
in the configuration $\bm{n}$
\begin{eqnarray}
\label{A}
A(\bm{n}) &\equiv& \sum_{(i,j)\in \Gamma} \sum_{\sigma=\ua\da} 
|\lambda_{ij\sigma}(\bm{n})|. 
\end{eqnarray}
The total number of jumps at time $t$ is 
$N_{t} = \sum_{(i,j)\in \Gamma}\sum_{\sigma=\ua\da} N^{t}_{ij \sigma}$.
By ordering the jumps according to the times $s_{k}$, $k=1,\dots, N_{t}$, 
at which they take place in the interval $[0,t)$,
we define a trajectory as the Markov chain
$\bm{n}_{1}, \bm{n}_2, \dots, \bm{n}_{N_{t}}$ 
generated from the initial configuration $\bm{n}_0$ 
by the stochastic dynamics described above.
Let us call 
$\lambda_{1},\lambda_{2}, \dots, \lambda_{N_{t}}$, 
$V_{1},V_2 \dots, V_{N_{t}}$ and
$A_{1},A_2 \dots, A_{N_{t}}$
the values of the matrix elements (\ref{lambda}), (\ref{potential})
and (\ref{A}) 
occurring along the trajectory, respectively.
For simplicity, we will indicate the last configuration 
reached after $N_t$ jumps as $\bm{n}_t = \bm{n}_{N_{t}}$.
We will also use the symbols $A_0=A(\bm{n}_0)$, $V_0=V(\bm{n}_0)$ and $s_0=0$.

According to Ref.~\cite{PRESILLA}, the following representation holds
\begin{eqnarray}
\label{TheFormulaa}
\langle \bm{n}|e^{-Ht} | \bm{n}_0\rangle &=&  \E  \left(
\delta_{ \bm{n} , \bm{n}_t} 
\mathcal{M}_{\bm{n}_0}^{[0,t)} \right), 
\end{eqnarray}
where the expectation $\E  \left( \cdot \right)$ has
to be taken with respect to the Poisson processes $N^{t}_{ij \sigma}$ and 
the stochastic functional $\mathcal{M}_{\bm{n}_0}^{[0,t)}$ is defined by
\begin{eqnarray}
\label{FORMULA CI}
\mathcal{M}_{\bm{n}_0}^{[0,t)}=e^{\int_{0}^{t}[A(\bm{n}_s)\eta-V(\bm{n}_s)]ds}.
\end{eqnarray}
The subscript $\bm{n}_0$ in $\mathcal{M}_{\bm{n}_0}^{[0,t)}$
specifies the initial state.
For real times an analogous representation holds
\begin{eqnarray}
\label{TheFormulaai}
\langle \bm{n}|e^{-iHt} | \bm{n}_0\rangle &=&  \E  \left(
\delta_{ \bm{n} , \bm{n}_t} 
\mathcal{M}_{\bm{n}_0}^{[0,it)} \right), 
\end{eqnarray}
where
\begin{eqnarray}
\label{FORMULA CIi}
\mathcal{M}_{\bm{n}_0}^{[0,it)}=i^{N_t}
e^{\int_{0}^{t}[A(\bm{n}_s)\eta-iV(\bm{n}_s)]ds}.
\end{eqnarray}
In the following, we will consider the case of imaginary times. 
Except when explicitly said,
all the formulas are trivially extended to the case of real times.
 
Any ground-state quantity can be obtained from the matrix 
elements (\ref{TheFormulaa}) by a proper manipulation and 
taking the limit $t \rightarrow \infty$. 
For instance, the ground-state energy is given by
\begin{eqnarray}
\label{E0} 
E_0 = \lim_{t \to \infty}
\frac{-\sum_{\bm{n}} \partial_t \langle \bm{n}|e^{-Ht} | \bm{n}_0\rangle}
{\sum_{\bm{n}} \langle \bm{n}|e^{-Ht} | \bm{n}_0\rangle }
= \lim_{t \to \infty} 
\frac{- \partial_t \E (\mathcal{M}_{\bm{n}_0}^{[0,t)})}
{\E (\mathcal{M}_{\bm{n}_0}^{[0,t)})}.
\end{eqnarray}
It is easy to see \cite{PRESILLA} that 
$- \partial_t \E (\mathcal{M}_{\bm{n}_0}^{[0,t)})=
\E (\mathcal{M}_{\bm{n}_0}^{[0,t)}\mathcal{H}(\bm{n}_t))$,
where $\mathcal{H}(\bm{n}_t)$ is given by
\begin{eqnarray} 
\label{Ht}
\mathcal{H}(\bm{n}_t)\equiv 
\sum_{\bm{n}'} \langle \bm{n}'|H| \bm{n}_t \rangle =
-[A(\bm{n}_t)\eta-V(\bm{n}_t)].
\end{eqnarray}
Equation (\ref{Ht}) is the local energy of the last visited configuration 
$\bm{n}_t$.
Therefore, Eq.~(\ref{E0}) becomes
\begin{eqnarray}
\label{E02}
E_0 &=& \lim_{t \to \infty} 
{\E (\mathcal{H}(\bm{n}_t)\mathcal{M}^{[0,t)}_{\bm{n}_0})
\over
\E (\mathcal{M}^{[0,t)}_{\bm{n}_0})} . 
\end{eqnarray}
These identities are valid if the initial configuration $\bm{n}_0$ 
is such that $ \langle E _0 | \bm{n}_0\rangle \neq 0 $. 
For a finite $t$, this scheme allows a good estimate of $E_{0}$ if 
$ t \gg (E_{1}-E_{0})^{-1} $, where $E_{1}$ is the first-excited state of $H$. 
This implies that $t$ must be increased by increasing the lattice size
$|\Lambda|$.

\section{Correlation functions}
\label{Correlation functions}
Let us consider a generic operator ${O}$. 
By using twice the Fock representation
of the identity operator and twice Eq.~(\ref{TheFormulaa}) with 
functionals $\mathcal{M}_{\bm{n}_0}^{[0,t)}$ 
and $\mathcal{M}'^{[0,t')}_{\bm{n}'_0}$,
respectively defined by two sets
of independent Poisson processes
$\{ N^{t}_{ij \sigma} \}$ and $\{ N'^{t'}_{ij \sigma} \}$, we have
\begin{eqnarray}\fl
\langle \bm{n}|e^{-Ht'} {O} e^{-Ht} |\bm{n}_0\rangle &=&
\sum_{\bm{n}_0'}\sum_{\bm{n}''} 
\langle \bm{n}|e^{-Ht'} |\bm{n}_0'\rangle
\langle \bm{n}_0'| {O} |\bm{n}''\rangle
\langle \bm{n}''|e^{-Ht} |\bm{n}_0\rangle 
\nonumber \\ &=&
\sum_{\bm{n}_0'}\sum_{\bm{n}''} 
\E  \left( \mathcal{M}'^{[0,t')}_{\bm{n}_0'} 
~\delta_{ \bm{n}'_{t'}, \bm{n}} 
~\langle\bm{n}_0'|O|\bm{n}''\rangle 
~\mathcal{M}_{\bm{n}_0}^{[0,t)}
~\delta_{ \bm{n}_t, \bm{n}''} 
\right)
\nonumber \\ &=&
\sum_{\bm{n}_0'} 
\E  \left( \delta_{ \bm{n}'_{t'}, \bm{n}}
~\mathcal{M}'^{[0,t')}_{\bm{n}_0'} 
~\langle\bm{n}_0'|O|\bm{n}_t\rangle 
~\mathcal{M}_{\bm{n}_0}^{[0,t)} 
\right),
\end{eqnarray}
where $\bm{n}'_{t'}$ is the configuration reached at time $t'$ starting 
from $\bm{n}_0'$. 
From this expression, we get
\begin{eqnarray}
\sum_{\bm{n}} \langle \bm{n}|e^{-Ht'} {O} e^{-Ht} |\bm{n}_0\rangle=
\sum_{\bm{n}}
\E  \left( \mathcal{M}'^{[0,t')}_{\bm{n}}
\langle \bm{n}|O|\bm{n}_t \rangle 
~\mathcal{M}_{\bm{n}_0}^{[0,t)}  \right).
\end{eqnarray}
The ground-state quantum expectation of the operator ${O}$,
assuming $\langle E_0 | E_0 \rangle =1$, is, therefore,
\begin{eqnarray}
\label{CORR}
\langle E_0|{O}| E_0 \rangle= 
\lim_{t,~t' \rightarrow \infty}
\frac{\sum_{\bm{n}}
\E  \left( \mathcal{M}'^{[0,t')}_{\bm{n}}
\langle \bm{n}|O|\bm{n}_t \rangle 
~\mathcal{M}_{\bm{n}_0}^{[0,t)} \right)}
{\E  \left( \mathcal{M}_{\bm{n}_0}^{[0,t+t')}\right)}.
\end{eqnarray}

We can consider two basic cases for the operator ${O}$.
\subsection{Diagonal operators}
In this case, $\langle \bm{n}'|{O}|\bm{n} 
\rangle=\delta_{\bm{n}',\bm{n}}O(\bm{n})$ and Eq.~(\ref{CORR}) becomes
\begin{eqnarray}
\label{CORRD}
\langle E_0|{O}| E_0 \rangle=
\lim_{t,~t' \rightarrow \infty}
\frac{
\E  \left( \mathcal{M}'^{[0,t')}_{\bm{n}_t}
~O(\bm{n}_t) 
~\mathcal{M}_{\bm{n}_0}^{[0,t)} \right)}
{\E  \left( \mathcal{M}_{\bm{n}_0}^{[0,t+t')}\right)}.
\end{eqnarray}
Note that
$\E ( \mathcal{M}'^{[0,t')}_{\bm{n}_t}
~\mathcal{M}_{\bm{n}_0}^{[0,t)} )
=\E ( \mathcal{M}_{\bm{n}_0}^{[0,t+t')})$,
so that $\langle E_0|{O}| E_0 \rangle=1$ if $O$ is the identity operator,
whereas for a single realization of the stochastic functionals we have
$\mathcal{M}'^{[0,t')}_{\bm{n}_t}
~\mathcal{M}_{\bm{n}_0}^{[0,t)}
\neq \mathcal{M}_{\bm{n}_0}^{[0,t+t')}$.

\subsection{Off-diagonal operators}
In this case, ${O}$ is typically 
given in terms of elementary operators ${O}_{ij\sigma}$ connecting
two different Fock states like
$\langle \bm{n}'| O_{ij\sigma} |\bm{n} \rangle=
O_{ij\sigma}(\bm{n})
\delta_{\bm{n}',\bm{n}^{i\sigma  \leftrightarrow j\sigma}}$, where 
$\bm{n}^{i\sigma  \leftrightarrow j\sigma}$ is the configuration obtained from
$\bm{n}$ exchanging $n_{i\sigma}$ with $n_{j\sigma}$.
Therefore, one has
\begin{eqnarray}
\label{CORROFF}
\langle E_0| O_{ij\sigma} | E_0 \rangle=
\lim_{t,~t' \rightarrow \infty}
\frac{
\E  \left( \mathcal{M}'^{[0,t')}_{\bm{n}_t^{i\sigma  \leftrightarrow j\sigma}}
~O_{ij\sigma}(\bm{n}_t) 
~\mathcal{M}_{\bm{n}_0}^{[0,t)} \right)}
{\E  \left( \mathcal{M}_{\bm{n}_0}^{[0,t+t')}\right)}.
\end{eqnarray}
Similar expressions hold for other off-diagonal operators connecting 
two generic Fock states.

\section{EPRMC algorithm}
\label{EPRMC algorithm}

\subsection{Pure sampling}
Equations (\ref{TheFormulaa}) and (\ref{FORMULA CI}) lend themselves 
to a statistical evaluation of the matrix elements
$\sum_{\bm{n}'} \langle \bm{n}'|e^{-Ht} | \bm{n}\rangle$ 
via a random sampling of jump times and trajectories.
As explained in Ref.~\cite{PRESILLA}, the practical algorithm works as follows.
We start by determining the active links in the initial configuration 
$\bm{n}_0$ assigned at time $0$ and make an extraction with uniform 
distribution to decide which of them jumps first,
say the link $(i_1j_1 \sigma_1)$.
We then extract the jump time $s_1$ according to the conditional probability
density 
\begin{eqnarray}
p_{A_0}(s)=A_0 \eta \exp(- A_0 \eta s),
\end{eqnarray} 
where $A_0$ is the number of active links before the first jump takes place.
The contribution to $\mathcal{M}_{\bm{n}_0}^{[0,t)}$ 
at the time of the first jump is, therefore,
\begin{eqnarray}
e^{(A_0\eta -V_0)s_1 }\theta(t-s_1)
+ e^{(A_0\eta-V_0)t}\theta(s_1-t).
\end{eqnarray}
According to Eq.~(\ref{FORMULA CI}), 
the contribution of a given trajectory is then obtained by multiplying the
factors corresponding to the different jumps determined in
an analogous way until the last jump takes place later than $t$,
\textit{i.e.},
\begin{eqnarray}
\label{FORMULA D}
{\cal M}^{[0,t)}_{\bm{n}_0} =
\left( 
\prod_{k=1}^{N_{t}} 
e^{(A_{k-1}\eta -V_{k-1})(s_{k}-s_{k-1})} 
\right) 
e^{(A_{N_{t}}\eta -V_{N_{t}})(t-s_{N_{t}})}
\end{eqnarray}
if $N_t > 0$, and 
${\cal M}^{[0,t)}_{\bm{n}_0} = e^{(A_0\eta-V_0) t}$ if $N_t=0$.

Let us consider $N$ independent trajectories obtained as described above 
and let $\mathcal{M}^{[0,t)(i)}_{\bm{n}_0}$ be the functional value 
(\ref{FORMULA CI}) calculated along the $i$-th trajectory.
From the law of large numbers we have
\begin{eqnarray}
\label{FORMULA CI1}
\E  \left(\mathcal{M}^{[0,t)}_{\bm{n}_0} \right)=
\lim_{N \rightarrow \infty}
\frac{1}{N} \sum_{i=1}^{N} \mathcal{M}_{\bm{n}_0}^{[0,t)(i)}.
\end{eqnarray}

\subsection{\label{SECRECONF}Reconfigurations}

Equation (\ref{FORMULA CI}) represents a product of 
$N_{t}$ random factors and, since $N_{t}$ grows with $t$, 
the fluctuations of the functional $\mathcal{M}_{\bm{n}_0}^{[0,t)}$ 
grow exponentially with $t$.
This implies that the number of trajectories
needed to have good statistical averages grows exponentially with $t$.
A similar problem has been successfully tackled some years ago 
in the framework of GFMC by the reconfiguration technique 
\cite{HETHERINGTON,CALANDRASORELLA}.
This technique can be adapted also to the present probabilistic representation.
In fact, for boson systems at imaginary times the  stochastic functional 
$\mathcal{M}_{\bm{n}_0}^{[0,t)}$ is always positive and can be thought
as a weight. 
Let us divide the time interval $[0,t)$ in $R$ subintervals of the same
length $\Delta t=t/R$. 
Let us label the times corresponding to the end-points of these intervals as
\begin{eqnarray}
t_{r}\equiv r\Delta t, \qquad r=0,\dots,R
\end{eqnarray}
and let $\bm{n}_{t_r}$ be the configuration reached at 
the time $t_r+0^+$ trough the dynamics 
described in Section \ref{Probabilistic representation}
(we recall that the Poisson processes are left continuous defined).
The following obvious identity follows from Eq.~(\ref{FORMULA CI})
\begin{eqnarray}
\label{PROD}
\mathcal{M}^{[0,t)}_{\bm{n}_0}=
\prod_{r=1}^{R}\mathcal{M}^{[t_{r-1},t_r)}_{\bm{n}_{t_{r-1}}},
\end{eqnarray}
which implies
\begin{eqnarray}
\label{FORMULA CIPROD}
\E  \left(\mathcal{M}^{[0,t)}_{\bm{n}_{0}} \right)=
\E  \left(\prod_{r=1}^{R}\mathcal{M}^{[t_{r-1},t_r)}_{\bm{n}_{t_{r-1}}}\right).
\end{eqnarray}
The functional $\mathcal{M}^{[t_{r-1},t_r)}_{\bm{n}_{t_{r-1}}}$ 
will be referred to as local weight. 

Essentially, the idea of the reconfiguration technique is the following.
Instead of extracting independent trajectories,
one carries on an ensemble of $M$ trajectories simultaneously
in order to perform dynamically, at the times $t_r$, 
a suitable replica of those with large weights, 
eliminating at the same time the others.
This replication/elimination of trajectories, 
also referred to as reconfiguration,
has to be done in such a way that the number of trajectories $M$ 
remains constant. 
At the end, one can substitute the average of
$\prod_{r=1}^{R}\mathcal{M}^{[t_{r-1},t_r)}_{\bm{n}_{t_{r-1}}}$
with the average of 
$\prod_{r=1}^{R}
\media{\mathcal{M}^{[t_{r-1},t_r)}_{\widetilde{\bm{n}}_{t_{r-1}}}}$,
where with $\bm{n}_{t_r} \rightarrow \widetilde{\bm{n}}_{t_r}$ 
we indicate the reconfiguration action at the time $t_r$, 
and with 
$\media{\mathcal{M}^{[t_{r-1},t_r)}_{\widetilde{\bm{n}}_{t_{r-1}}}}$ 
the uniform ``average'' of the local weights over the 
$M$ reconfigured trajectories 
(we use quotation marks since this quantity is itself a random variable). 
Hence, the remarkable advantage of using the reconfigurations
is that, if the functional 
$\prod_{r=1}^{R}\mathcal{M}^{[t_{r-1},t_r)}_{\bm{n}_{t_{r-1}}}$
has variance $\Delta^{R^*}$, the variance of
$\prod_{r=1}^{R}
\media{\mathcal{M}^{[t_{r-1},t_r)}_{\widetilde{\bm{n}}_{t_{r-1}}}}$
will be roughly $(\Delta/\sqrt{M})^{R^*}$, 
where $\Delta$ is the variance of the local weights and $R^*<R$ is the number
of subintervals in which the local weights become approximately independent.

\subsubsection{Reconfiguration algorithm.}\label{SECRECALG}
here, we describe in detail the reconfiguration algorithm at imaginary times
postponing the relative proof to the next Section.
We will indicate with 
$\bm{n}_{t_{r}}^{(i)}$, $r=0,\dots,R$, and 
$\mathcal{M}^{[t_{r-1},t_r)(i)}_{\bm{n}_{t_{r-1}}^{(i)}}$, $r=1,\dots,R$,
the configurations and the local weights 
of the $i$-th trajectory and define the corresponding $M$-component vectors 
as $\underline{\bm{n}_{t_r}}$ and 
$\underline{\mathcal{M}^{[t_{r-1},t_r)}_{\bm{n}_{t_{r-1}}}}$,
respectively.
We shall use also the operator symbols $\mathcal{D}$ and $\mathcal{R}$: 
$\mathcal{D}$ applied to 
the configurations $\underline{\bm{n}_{t_r}}$ gives the configurations
$\underline{\bm{n}_{t_{r+1}}}$ according to the dynamics
defined in Section \ref{Probabilistic representation} 
along the time interval $[t_r,t_{r+1})$, whereas
$\mathcal{R}$ gives the reconfigured configurations 
$\underline{\widetilde{\bm{n}}_{t_r}}=\mathcal{R} \underline{\bm{n}_{t_r}}$.

First step.
Define 
$\underline{\widetilde{\bm{n}}_{t_0}} = \underline{\bm{n}_{t_0}}$
with $\bm{n}_{t_0}^{(i)} = \bm{n}_{0}$ for $i=1,\ldots,M$. 
At the initial time $t_0=0$, all the $M$ 
trajectories starting from the initial configuration
$\bm{n}_0$ follow the dynamics $\mathcal{D}$ and reach
the configurations 
$\underline{\bm{n}_{t_1}}=\mathcal{D} \underline{\bm{n}_{t_0}}$.
Correspondingly, evaluate the $M$ local weights along the time interval
$[0,t_1)$, 
$\underline{\mathcal{M}^{[0,t_1)}_{\widetilde{\bm{n}}_{t_0}}}$,
and compute their average
\begin{eqnarray}
\media{\mathcal{M}^{[0,t_1)}_{\widetilde{\bm{n}}_{t_0}}}\equiv 
\frac{1}{M} \sum_{l=1}^{M}\mathcal{M}^{[0,t_1)(l)}_{\widetilde{\bm{n}}_{t_0}}.
\end{eqnarray}

Second step.
Perform the reconfiguration 
$\underline{\bm{n}_{t_1}} \rightarrow \underline{\widetilde{\bm{n}}_{t_1}}
=\mathcal{R} \underline{\bm{n}_{t_1}}$.
The new configurations are obtained by drawing out them randomly from 
the old ones, $\underline{\bm{n}_{t_1}}$, 
according to the probabilities
\begin{eqnarray}
\label{RECONFIGURATION1}
\mathcal{P}_{t_1}^{(i)}\equiv
\frac{\mathcal{M}^{[0,t_1)(i)}_{\widetilde{\bm{n}}_{t_0}}}{\sum_{l=1}^{M} 
\mathcal{M}^{[0,t_1)(l)}_{\widetilde{\bm{n}}_{t_0}}}.
\end{eqnarray}
The new configurations $\underline{\widetilde{\bm{n}}_{t_1}}$ are used 
as starting configurations of the $M$ trajectories for the time
interval $[t_1,t_2)$ and, through the dynamics $\mathcal{D}$,
are mapped into $\mathcal{D} \underline{\widetilde{\bm{n}}_{t_1}}$.
Correspondingly, evaluate the $M$ local weights
$\underline{\mathcal{M}^{[t_1,t_2)}_{\widetilde{\bm{n}}_{t_1}}}$
and compute their average
\begin{eqnarray}
\media{\mathcal{M}^{[t_1,t_2)}_{\widetilde{\bm{n}}_{t_1}}}\equiv 
\frac{1}{M} 
\sum_{l=1}^{M}\mathcal{M}^{[t_1,t_2)(l)}_{\widetilde{\bm{n}}_{t_1}^{(l)}}.
\end{eqnarray}

Third step.
Perform the reconfiguration
$\mathcal{D} \underline{\widetilde{\bm{n}}_{t_1}} 
\rightarrow \underline{\widetilde{\bm{n}}_{t_2}} =
\mathcal{R}\mathcal{D} \underline{\widetilde{\bm{n}}_{t_1}}$
by drawing out the new configurations randomly
from the old ones according to the probabilities
\begin{eqnarray}
\label{RECONFIGURATION2}
\mathcal{P}_{t_2}^{(i)}\equiv
\frac{\mathcal{M}^{[t_1,t_2)(i)}_{\widetilde{\bm{n}}_{t_1}^{(i)}}}
{\sum_{l=1}^{M} 
\mathcal{M}^{[t_1,t_2)(l)}_{\widetilde{\bm{n}}_{t_1}^{(l)}}}.
\end{eqnarray}
The new configurations $\underline{\widetilde{\bm{n}}_{t_2}}$ are used 
as starting configurations in the time interval $[t_2,t_3)$.
Evaluate the local weights 
$\underline{\mathcal{M}^{[t_2,t_3)}_{\widetilde{\bm{n}}_{t_2}}}$
and compute their average
\begin{eqnarray}
\media{\mathcal{M}^{[t_2,t_3)}_{\widetilde{\bm{n}}_{t_2}}}\equiv 
\frac{1}{M}
\sum_{l=1}^{M}\mathcal{M}^{[t_2,t_3)(l)}_{\widetilde{\bm{n}}_{t_2}^{(l)}}.
\end{eqnarray}

By iterating this procedure for $R$ steps, we arrive to 
the final configurations 
$\mathcal{D} \underline{\widetilde{\bm{n}}_{t_{R-1}}} 
=\mathcal{D} (\mathcal{RD})^{R-1} \underline{\bm{n}_{t_0}}$,
with $R$ computed averages 
$\media{\mathcal{M}^{[t_{r-1},t_r)}_{\widetilde{\bm{n}}_{t_{r-1}}}}$, 
$r=1,\dots,R$.
As we will prove later, the following identity holds
\begin{eqnarray}
\label{RECONFIGURATION3}
\E \left(\mathcal{M}^{[0,t)}_{\bm{n}_0}\right)=
\widetilde{\E} 
\left(\prod_{r=1}^{R}
\media{\mathcal{M}^{[t_{r-1},t_r)}_{\widetilde{\bm{n}}_{t_{r-1}}}}\right),
\end{eqnarray}
where $\widetilde{\E}$ indicates the expectation 
in which the configurations $\underline{\widetilde{\bm{n}}_{t_r}}$ 
are obtained by the reconfiguration procedure described above.  
Explicitly, Eq.~(\ref{RECONFIGURATION3})
implies that to evaluate the expectation 
$\E ( \mathcal{M}^{[0,t)}_{\bm{n}_0} )$,
instead of Eq.~(\ref{FORMULA CI1}), we can use
\begin{eqnarray}
\E \left(\mathcal{M}^{[0,t)}_{\bm{n}_0}\right)=
\lim_{M \rightarrow \infty}
\prod_{r=1}^{R}
\media{\mathcal{M}^{[t_{r-1},t_r)}_{\widetilde{\bm{n}}_{t_{r-1}}}},
\end{eqnarray}
or, more generally, simulating $N$ independent samples each one
composed by $M$ reconfigured trajectories,
\begin{eqnarray}
\label{RECONFIGURATION3.1}
\E \left(\mathcal{M}^{[0,t)}_{\bm{n}_0}\right)=
\lim_{MN \rightarrow \infty} \frac{1}{N}
\sum_{p=1}^{N}\prod_{r=1}^{R}
\media{\mathcal{M}^{[t_{r-1},t_r)}_{\widetilde{\bm{n}}_{t_{r-1}}}}^{(p)}.
\end{eqnarray}
The label $(p)$ in Eq.~(\ref{RECONFIGURATION3.1}) means $p$-th sample.
Note that for $M=1$ we recover Eq.~(\ref{FORMULA CI1}).

All what we said about the functional $\mathcal{M}^{[0,t)}_{\bm{n}_0}$ 
can be repeated for the functional
$\mathcal{M}^{[0,t)}_{\bm{n}_0}\delta_{ \bm{n} , \bm{n}_t}$. 
In this case, Eq.~(\ref{RECONFIGURATION3}) becomes
\begin{eqnarray}
\label{RECONFIGURATION4}
\E \left(\mathcal{M}^{[0,t)}_{\bm{n}_0}\delta_{ \bm{n} , \bm{n}_t}\right)=
\widetilde{\E} \left(\prod_{r=1}^{R-1}
\media{\mathcal{M}^{[t_{r-1},t_r)}_{\widetilde{\bm{n}}_{t_{r-1}}}} 
\frac{1}{M}
\sum_{l=1}^M\mathcal{M}^{[t_{R-1},t)(l)}_{\widetilde{\bm{n}}_{t_{R-1}}^{(l)}} 
\delta_{ \bm{n} , (\mathcal{D}\underline{\widetilde{\bm{n}}_{t_{R-1}}})^{(l)} }\right).
\end{eqnarray}
Equation (\ref{RECONFIGURATION4}) allows to calculate the numerator 
of Eq.~(\ref{E02}) as
\begin{eqnarray}
\label{RECONFIGURATION5}\fl
\E \left(\mathcal{M}^{[0,t)}_{\bm{n}_0}\mathcal{H}(\bm{n}_t)\right)=
\widetilde{\E} \left(\prod_{r=1}^{R-1}
\media{\mathcal{M}^{[t_{r-1},t_r)}_{\widetilde{\bm{n}}_{t_{r-1}}}} 
\frac{1}{M}
\sum_{l=1}^M\mathcal{M}^{[t_{R-1},t)(l)}_{\widetilde{\bm{n}}_{t_{R-1}}^{(l)}} 
\mathcal{H}((\mathcal{D} \underline{\widetilde{\bm{n}}_{t_{R-1}}})^{(l)}) 
\right).
\end{eqnarray}

\subsubsection{Correlation functions.}
Let us now consider the reconfiguration procedure for the functionals
introduced in Eqs.~(\ref{CORRD}) and (\ref{CORROFF}) to obtain 
the correlation functions.
In this case, we perform $R$ steps in the first interval $[0,t)$ 
and $R'$ steps in the second interval $[0,t')$.
All the quantities relative to the second interval $[0,t')$ will be
be indicated with a prime. 
In the pure sampling, the initial configurations
of the second interval $[0,t')$ are equal to the final ones of the first
interval $[0,t)$: $\underline{\bm{n}'_{t'_0}}=\underline{\bm{n}_{t_{R}}}$.
For diagonal operators, we have
\begin{eqnarray}\fl
\label{RCORRD}
\E  \left(\mathcal{M}'^{[0,t')}_{\bm{n}_t} 
~O(\bm{n}_t) ~\mathcal{M}_{\bm{n}_0}^{[0,t)}\right) &=& 
\widetilde{\E} \left(
\prod_{r=1}^{R}
\media{\mathcal{M}^{[t_{r-1},t_r)}_{\widetilde{\bm{n}}_{t_{r-1}}}} 
\prod_{r'=1}^{R'-1}
\media{\mathcal{M}'^{[t'_{r'-1},t'_{r'})}_{\widetilde{\bm{n}}'_{t'_{r'-1}}}} 
\right. \nonumber \\ && \times \left.
\frac{1}{M}\sum_{l=1}^M
\mathcal{M}'^{[t'_{R'-1},t')(l)}_{\widetilde{\bm{n}}'^{(l)}_{t_{R'-1}}} 
O 
((\mathcal{R}^{R'}\mathcal{D}\underline{\widetilde{\bm{n}}_{t_{R-1}}})^{(l)})
\right),
\end{eqnarray}
where now the configurations 
$\mathcal{R}^{R'}\mathcal{D} \underline{\widetilde{\bm{n}}_{t_{R-1}}}$ 
are obtained by updating the intermediate configurations 
at time $t_R$, namely 
$\mathcal{D} \underline{\widetilde{\bm{n}}_{t_{R-1}}}$,
$R'$ times according to the successive $R'$ steps.
For off-diagonal operators, we have
\begin{eqnarray}\fl
\label{RCORROFF}
\E \left( \mathcal{M}'^{[0,t')}_{\bm{n}_t^{i\sigma  \leftrightarrow j\sigma}}
~O_{ij\sigma}(\bm{n}_t) 
~\mathcal{M}_{\bm{n}_0}^{[0,t)}\right) &=& 
\widetilde{\E} \left(
\prod_{r=1}^{R}
\media{\mathcal{M}^{[t_{r-1},t_r)}_{\widetilde{\bm{n}}_{t_{r-1}}}} 
\prod_{r'=1}^{R'-1}
\media{
\mathcal{M}'^{[t'_{r'-1},t'_{r'})}_{\widetilde{\bm{n}}^{\mathrm{ex}}_{t'_{r'-1}}}}
\right. \nonumber \\ && \times \left.
\frac{1}{M}\sum_{l=1}^M
\mathcal{M}'^{[t'_{R'-1},t')(l)}_{\widetilde{\bm{n}}^{\mathrm{ex}(l)}_{t'_{R'-1}} } 
O_{ij\sigma}
((\mathcal{R}^{R'}\mathcal{D}\underline{\widetilde{\bm{n}}_{t_{R-1}}})^{(l)})
\right),
\end{eqnarray}
where
$\underline{\widetilde{\bm{n}}^\mathrm{ex}_{t'_{r'-1}}}$, $r'=1,\dots, R'$, 
are the configurations obtained after $r'$ steps starting from the intermediate
configurations 
$(\underline{\mathcal{D}\widetilde{\bm{n}}_{t_{R-1}} 
})^{i\sigma \leftrightarrow j\sigma} $ 
in which the occupations of sites $i$ and $j$ with spin $\sigma$ 
have been exchanged, \textit{i.e.} 
$\underline{\widetilde{\bm{n}}^\mathrm{ex}_{t'_{r'-1}}}
=(\mathcal{R}\mathcal{D})^{r'-1}
( \underline{\mathcal{D}\widetilde{\bm{n}}_{t_{R-1}}} 
)^{i\sigma \leftrightarrow j\sigma}$.

\subsubsection{Real times.}
A reconfiguration procedure can be performed also at real times.
In this case, the stochastic functional $\mathcal{M}^{[0,it)}_{\bm{n}_0}$
is complex and we separate the contributions from the $R$ 
time intervals in their moduli and arguments, \textit{i.e.}
\begin{eqnarray}
\mathcal{M}^{[it_{r-1},it_r)}_{\bm{n}_{t_{r-1}}}=
| \mathcal{M}^{[it_{r-1},it_r)}_{\bm{n}_{t_{r-1}}} |
~
e^{ i 
\Phi^{[t_{r-1},t_r)}_{\bm{n}_{t_{r-1}}} },
\end{eqnarray}
where
\begin{eqnarray}
\label{weighti}
| \mathcal{M}^{[it_{r-1},it_r)}_{\bm{n}_{t_{r-1}}} | = 
e^{\int_{t_{r-1}}^{t_r} A(\bm{n}_s)\eta ds},
\\
\Phi^{[t_{r-1},t_r)}_{\bm{n}_{t_{r-1}}} = 
\frac{\pi}{2}(N_{t_r}-N_{t_{r-1}}) - 
\int_{t_{r-1}}^{t_r}V(\bm{n}_s)ds.
\end{eqnarray}
The moduli can be used as local weights for the reconfiguration
operator $\mathcal{R}$.
All the steps described in Section (\ref{SECRECALG})
remain unchanged except for the last factor, which takes into account 
the $R$ phase factors reconstructing the original stochastic functional.
The final result result is
\begin{eqnarray}\fl
\label{RECONFIGURATIONi}
\E \left(\mathcal{M}^{[0,it)}_{\bm{n}_{0}}\right)=&&
\widetilde{\E} 
\left(\prod_{r=1}^{R-1}
\media{| \mathcal{M}^{[it_{r-1},it_r)}_{\widetilde{\bm{n}}_{t_{r-1}}}| }
~\frac{1}{M}\sum_{l=1}^M
|\mathcal{M}^{[it_{R-1},it_R)(l)}_{\widetilde{\bm{n}}_{t_{R-1}}^{(l)}}|
~e^{ i 
\sum_{r=1}^{R}
\Phi^{[t_{r-1},t_r)(l)}_{(\mathcal{R}^{R-r}
\underline{\widetilde{\bm{n}}_{t_{r-1}}})^{(l)}} }
\right).
\end{eqnarray}

\subsection{Importance sampling}
\label{Importance sampling}
Although the reconfiguration method controls
the growth of the fluctuations of $\mathcal{M}^{[0,t)}$
along the trajectories, since the dimension of the Fock space 
grows exponentially with the lattice size, an extraction of
the jumping links by importance sampling also may 
be mandatory to reduce the statistical errors
of the local weights \cite{CEPERLEYKALOS}.
If some \textit{a priori} approximation $|g\rangle$ 
of the ground state is known, which has the property
$\langle \bm{n}| g \rangle \in \mathbb{R} \setminus 0$
for any Fock state $|\bm{n} \rangle$, then instead to
sample directly the operator $\exp(-Ht)$, it can be notably advantageous
to sample the isospectral operator $\exp(-H_gt)$,
where $\langle \bm{n}'|H_g|\bm{n}\rangle\equiv 
\langle \bm{n}'|g\rangle\langle\bm{n}'|H|\bm{n}\rangle
{\langle \bm{n}|g\rangle}^{-1}$.  

As explained in Refs.~\cite{PRESILLA} and \cite{PRESILLA1},
if $|g\rangle$ 
is a guiding function in the sense specified above,
the generalization of the present algorithm
to the case with importance sampling consists
in replacing the number of active links, 
$A(\bm{n}) \equiv \sum_{(i,j)\in \Gamma}\sum_{\sigma=\ua\da} 
|\lambda_{ij\sigma}(\bm{n})|$, 
in all the previous formulas with the quantity 
\begin{eqnarray}
\label{Ag}
A_g(\bm{n}) &\equiv& 
\sum_{(i,j)\in \Gamma} \sum_{\sigma=\ua\da} 
\left|\lambda_{ij\sigma}(\bm{n})\frac
{\langle \bm{n}\oplus \bm{1}_{i\sigma} \oplus \bm{1}_{j\sigma}| g \rangle}
{\langle \bm{n}| g \rangle}\right|.
\end{eqnarray}
Correspondingly, the probability density for the jump times becomes
\begin{eqnarray}
\label{pAg}
p_{A_g}(s)=A_g \eta \exp(- A_g \eta s),
\end{eqnarray} 
and the extraction of the jumping link $(i,j,\sigma)$
among the active ones 
must be performed according to the probabilities
$|\langle \bm{n}\oplus \bm{1}_{i\sigma} \oplus \bm{1}_{j\sigma}| g \rangle
{\langle \bm{n}| g \rangle}^{-1}|/A_g(\bm{n})$.
Finally, the stochastic functional (\ref{FORMULA CI}) is modified as 
\begin{eqnarray}
\label{MSF}
\mathcal{M}_{g,\bm{n}_0}^{[0,t)}=
e^{\int_{0}^{t}[A_g(\bm{n}_s)\eta-V(\bm{n}_s)]ds}.
\end{eqnarray}

The advantage of using importance sampling becomes
clear considering the local energy associated to $H_g$
\begin{eqnarray}
\label{Htg}
\mathcal{H}_g(\bm{n}_t)\equiv 
\sum_{\bm{n}'}
\langle \bm{n}'|g\rangle\langle\bm{n}'|H| \bm{n}_t \rangle
{\langle\bm{n}_t|g\rangle}^{-1} =
-[A_g(\bm{n}_t)\eta-V(\bm{n}_t)].
\end{eqnarray}
In fact, in the limit $|g\rangle \rightarrow |E_0\rangle$ one has
$\mathcal{H}_g(\bm{n}_t) \rightarrow E_0$ and accordingly 
${\cal M}_{g,\bm{n}_0}^{[0,t)} \rightarrow \exp(-E_0t)$
so that the fluctuations vanish.

For any choice of the guiding function $|g\rangle$, the modified 
stochastic functional (\ref{MSF}) provides unbiased representations
of the ground-state energy $E_0$ as well as of the expectation of
a generic operator $O$ in the ground state $|E_0\rangle$ of $H$. 
In fact, Eq.~(\ref{E02}) now reads 
\begin{eqnarray}
\lim_{t \to \infty} 
{\E (\mathcal{H}_g(\bm{n}_t)\mathcal{M}^{[0,t)}_{g,\bm{n}_0})
\over
\E (\mathcal{M}^{[0,t)}_{g,\bm{n}_0})} 
= E_{0g},
\end{eqnarray}
where $E_{0g}$ is the ground-state energy of $H_g$, which, however,  
is an operator isospectral to $H$.
On the other hand, Eq.~(\ref{CORR}) written in terms of a 
$g$-modified operator $O_g$ becomes
\begin{eqnarray}
\lim_{t,~t' \rightarrow \infty}
\frac{\sum_{\bm{n}}
\E  \left( \mathcal{M}'^{[0,t')}_{g,\bm{n}}
\langle \bm{n}|O_g|\bm{n}_t \rangle 
~\mathcal{M}_{g,\bm{n}_0}^{[0,t)} \right)}
{\E  \left( \mathcal{M}_{g,\bm{n}_0}^{[0,t+t')}\right)}
= \langle E_{0g}|O_g| E_{0g} \rangle.
\end{eqnarray}
By using 
$\langle \bm{n} | E_{0g} \rangle = \langle \bm{n} | g \rangle 
\langle \bm{n} | E_{0} \rangle$ 
and
$\langle E_{0g} | \bm{n} \rangle = \langle \bm{n} | g \rangle^{-1} 
\langle E_{0} | \bm{n} \rangle$,
it is simple to see that 
$\langle E_{0g}|O_g| E_{0g} \rangle = \langle E_0|{O}| E_0 \rangle$
if we choose $O_g$ as the operator defined by 
$\langle \bm{n}'|O_g|\bm{n}\rangle\equiv 
\langle \bm{n}'|g\rangle\langle\bm{n}'|O|\bm{n}\rangle
{\langle \bm{n}|g\rangle}^{-1}$.  
Note that, in the case of diagonal operators, $O_g=O$.

Importance sampling may be useful also for a different purpose, 
namely the determination of the transition amplitudes
$\langle g |e^{-iHt} | \bm{n}_0\rangle$ between two chosen
states $|\bm{n}_0\rangle$ and $|g\rangle$. 
This is particularly interesting at real times and we illustrate the 
idea in this case.
If $|g\rangle$ is a generic state with the property
$\langle \bm{n}| g \rangle \in \mathbb{R} \setminus 0$
so that the isospectral Hamiltonian $H_g$ is well defined, 
we have
\begin{eqnarray}
\label{realimportance}
\sum_{\bm{n}}
\langle \bm{n}|e^{-iH_gt} | \bm{n}_0\rangle = 
\langle g | \bm{n}_0\rangle  
\langle g |e^{-iHt} | \bm{n}_0 \rangle. 
\end{eqnarray}
Since the expectation of the stochastic functional 
$\mathcal{M}_{\bm{n}_0}^{[0,it)}$ with the modified rules 
(\ref{Ag}) and (\ref{pAg}) provides an exact representation of the l.h.s.
of Eq.~(\ref{realimportance}), we obtain 
the transition amplitudes $\langle g |e^{-iHt} | \bm{n}_0\rangle$ 
up to the constant $\langle g | \bm{n}_0\rangle$.

\section{Proof of the reconfiguration algorithm}
\label{Proof of the reconfiguration method}
In this Section, we prove Eqs.~(\ref{RECONFIGURATION3}-\ref{RCORROFF}).
Let us consider an ensemble of $M$ simultaneous trajectories
obtained by the dynamics described in Section 
\ref{Probabilistic representation}
starting from the initial configuration $\bm{n}_0$. 
Let 
$P_R(\underline{\mathcal{M}^{[t_0,t_1)}_{\bm{n}_{t_0}}},
\underline{\mathcal{M}^{[t_1,t_2)}_{\bm{n}_{t_1}}},
\ldots,\underline{\mathcal{M}^{[t_{R-1},t)}_{\bm{n}_{t_{R-1}}}};
\underline{\bm{n}_{t_0}},\underline{\bm{n}_{t_1}},
,\ldots,\underline{\bm{n}_{t_R}})$ be
the probability density to have a realization in which the $M$ 
trajectories have local weights
$\underline{\mathcal{M}^{[t_0,t_1)}_{\bm{n}_{t_0}}},
\underline{\mathcal{M}^{[t_1,t_2)}_{\bm{n}_{t_1}}},
\ldots,\underline{\mathcal{M}^{[t_{R-1},t)}_{\bm{n}_{t_{R-1}}}}$
and configurations $\underline{\bm{n}_{t_0}},\underline{\bm{n}_{t_1}},
,\ldots,\underline{\bm{n}_{t_R}}$ at the times 
$t_0,t_1,\ldots,t_R$, respectively.
For simplicity, here we shall often use 
$\underline{\mathcal{M}_{r-1}}$
for $\underline{\mathcal{M}_{\bm{n}_{t_{r-1}}}^{[t_{r-1},t_r)}}$
and $\underline{\bm{n}_{r}}$ for $\underline{\bm{n}_{t_r}}$.
Since the $M$ trajectories are independent, 
if we take $\bm{n}_0^{(l)}=\bm{n}_0$ for $l=1,\dots,M$, we have
\begin{eqnarray}\fl
\label{PROOF1}
\E  \left(
\mathcal{M}_{\bm{n}_0}^{[0,t)}
\delta_{ \bm{n} , \bm{n}_t} \right) &=&
\E  \left(
\prod_{r=1}^{R}\mathcal{M}_{r-1}
\delta_{ \bm{n} , \bm{n}_R} \right)
=
\E  \left(
\frac{1}{M}\sum_{l=1}^M\prod_{r=1}^{R}\mathcal{M}_{r-1}^{(l)}
\delta_{ \bm{n} , \bm{n}_R^{(l)}} \right). 
\end{eqnarray}
Consider, then, the following identity
\begin{eqnarray}\fl
\label{PROOF2}
\frac{1}{M}\sum_{l=1}^M\prod_{r=1}^{R}\mathcal{M}_{r-1}^{(l)}
\delta_{ \bm{n} , \bm{n}_R^{(l)}} &=& 
\left( \sum_{l=1}^M \mathcal{M}_0^{(l)}p_{0}^{(l)} \right)
\left( \sum_{l=1}^M \mathcal{M}_1^{(l)}p_{1}^{(l)} \right)              
\cdots
\left( \sum_{l=1}^M \mathcal{M}_{R-2}^{(l)}p_{R-2}^{(l)} \right)
 \nonumber \\ &&\times
\left( \sum_{l=1}^M 
\mathcal{M}_{R-1}^{(l)}\delta_{\bm{n},\bm{n}_R^{(l)}}p_{R-1}^{(l)} \right),
\end{eqnarray}
where the quantities 
$\underline{p_{0}},\underline{p_{1}},\ldots,\underline{p_{R-1}}$
are defined recursively by
\begin{eqnarray}
\label{p}
\left\{
\begin{array}{l}
p_0^{(i)}=\frac{1}{M} \\
p_r^{(i)}=\frac{\mathcal{M}_{r-1}^{(i)}p_{r-1}^{(i)}}
{\sum_{l=1}^M \mathcal{M}_{r-1}^{(l)}p_{r-1}^{(l)}}, 
\qquad r=1,\dots,R-1
\end{array}
\right. .
\end{eqnarray}
Equations (\ref{PROOF1}) and (\ref{PROOF2}) lead to 
\begin{eqnarray}
\label{PROOF3}
\E  \left(
\prod_{r=1}^{R}\mathcal{M}_{r-1}
\delta_{ \bm{n} , \bm{n}_R} \right)=
\E  \left(
\prod_{r=1}^{R-1}\media{\mathcal{M}_{r-1}}_{w}
\media{\mathcal{M}_{R-1}\delta_{ \bm{n} , \bm{n}_R}}_{w} \right),
\end{eqnarray}
where the weighted ``averages'', $\media{\mathcal{M}_r}_{w}$ and 
$\media{\mathcal{M}_{R-1}\delta_{ \bm{n} , \bm{n}_R}}_{w}$,
are defined as the weighted sums 
$\media{\mathcal{M}_r}_{w}=\sum_{l=1}^M \mathcal{M}_r^{(l)}p_{r}^{(l)}$
and $\media{\mathcal{M}_{R-1}\delta_{ \bm{n} , \bm{n}_R}}_{w}=
\sum_{l=1}^M \mathcal{M}_{R-1}^{(l)}\delta_{
\bm{n} , \bm{n}_R^{(l)}}p_{R-1}^{(l)}$,
respectively.

Up to now the quantities $\underline{p_{r}}$ 
have been thought as stochastic variables.
Actually, since the components $p_r^{(l)}$ 
are positive and normalized to 1, we can 
interpret them as probabilities to modify the original 
probability density $P_R$.
We introduce a new probability density $\widetilde{P}_R$ that, 
besides taking into account the dynamics $\mathcal{D}^R$, 
includes the probabilities $\underline{p_r}$, for $r=0,\ldots,R-1$.
In this case, if we indicate with $\underline{\widetilde{\bm{n}}_{0}},
\underline{\widetilde{\bm{n}}_{1}}
,\ldots,\underline{\widetilde{\bm{n}}_{R-1}},
\mathcal{D} \underline{\widetilde{\bm{n}}_{R-1}}$ 
the configurations extracted according to the  
probability density $\widetilde{P}_R$, 
Eq.~(\ref{PROOF3}) transforms into 
\begin{eqnarray}\fl
\label{PROOF4}
\E  \left(
\prod_{r=1}^{R}\mathcal{M}_{r-1}
\delta_{ \bm{n} , \bm{n}_R} \right)=
\widetilde{\E} \left( \prod_{r=1}^{R-1}
\media{\mathcal{M}_{\widetilde{\bm{n}}_{r-1}}^{[t_{r-1},t_r)}} 
\frac{1}{M}
\sum_{l=1}^M
\mathcal{M}_{\widetilde{\bm{n}}_{R-1}^{(l)}}^{[t_{R-1},t_R)(l)} 
\delta_{ \bm{n} , (\mathcal{D}\underline{\widetilde{\bm{n}}_{R-1}})^{(l)}  }
\right),
\end{eqnarray}
where $\widetilde{\E}( \cdot )$ means expectation with respect to 
 $\widetilde{P}_R$ and the weighted ``averages'' 
$\media{\mathcal{M}_r}_{w}$ have been
substituted by uniform ``averages'' over the new configurations,
$\media{\mathcal{M}_{\widetilde{\bm{n}}_{r-1}}^{[t_{r-1},t_r)}}= 
\sum_{l=1}^M 
\mathcal{M}_{\widetilde{\bm{n}}_{r-1}^{(l)}}^{[t_{r-1},t_r)(l)}/M$.

Equations  (\ref{PROOF1}) and (\ref{PROOF4}) reproduce
Eq.~(\ref{RECONFIGURATION4}). 
To conclude the proof, we still have to show that 
the algorithm described in Section \ref{SECRECALG} 
coincides with sampling the configurations 
$\underline{\widetilde{\bm{n}}_{0}},\underline{\widetilde{\bm{n}}_{1}}
,\ldots,\underline{\widetilde{\bm{n}}_{R-1}},
\mathcal{D} \underline{\widetilde{\bm{n}}_{R-1}} $ 
according to the probability density $\widetilde{P}_R$.
For $M$ trajectories with local weights $\mathcal{M}_{r-1}^{(i)}$,
let us define the following probabilities
\begin{equation}
\label{Pstorto}
\mathcal{P}_{r}^{(i)}=
\frac{\mathcal{M}_{r-1}^{(i)}}
{\sum_{l=1}^M \mathcal{M}_{r-1}^{(l)}},
\qquad r=1,\dots,R-1.
\end{equation} 
Due to the recursiveness of Eq.~(\ref{p}), for $r \geq 1$,
we have 
\begin{equation}
\label{p1}
p_r^{(i)} = C_r \prod_{r'=1}^{r}\mathcal{P}_{r'}^{(i)},
\end{equation} 
where $C_r$ is a normalization constant independent of the trajectory
index $(i)$.
This allows to realize the transformation 
$P_r \rightarrow \widetilde{P}_r$ recursively for $r=1, \ldots, R$. 
At the first step $r=1$,
since $\underline{p_0}$ is uniform we do not have to
reconfigure and $\underline{\widetilde{\bm{n}}_{0}}=
\underline{\bm{n}_0}$. 
The density $\widetilde{P}_1$  will be then sampled through the vectors 
$\underline{\widetilde{\bm{n}}_{0}}$ and 
$\mathcal{D} \underline{\widetilde{\bm{n}}_{0}}$. 
Suppose now to have sampled the density $\widetilde{P}_{r}$
through the vectors 
$\underline{\widetilde{\bm{n}}_{0}},\underline{\widetilde{\bm{n}}_{1}}
,\ldots,\underline{\widetilde{\bm{n}}_{r-1}},
\mathcal{D} \underline{\widetilde{\bm{n}}_{r-1}}$.
To sample the density $\widetilde{P}_{r+1}$ we must
change the arrival vector of configurations 
$\mathcal{D} \underline{\widetilde{\bm{n}}_{r-1}}$ 
into a new vector $\underline{\widetilde{\bm{n}}_r}$ 
according to the probabilities $\underline{\mathcal{P}_{r}}$,
with components
\begin{eqnarray}
\mathcal{P}_{r}^{(i)} =
\frac{\mathcal{M}^{[t_{r-1},t_r)(i)}_{\widetilde{\bm{n}}_{t_{r-1}}}}{\sum_{l=1}^{M} 
\mathcal{M}^{[t_{r-1},t_r)(l)}_{\widetilde{\bm{n}}_{t_{r-1}}}}.
\end{eqnarray}
With a further dynamic step we get 
$\mathcal{D} \underline{\widetilde{\bm{n}}_{r}}$.
The distribution $\widetilde{P}_R$ is sampled by iterating this procedure 
$R$ times.
This is exactly the procedure explained in Section \ref{SECRECALG} and
the reconfiguration algorithm is proved.

Equation (\ref{RECONFIGURATION3}) follows easily by summing 
Eq.~(\ref{RECONFIGURATION4}) over $\bm{n}$.
Finally, Eq.~(\ref{RECONFIGURATION5}) can be obtained 
multiplying $\mathcal{M}_{\bm{n}_0}^{[0,t)}\delta_{\bm{n},\bm{n}_t}$ 
by $\mathcal{H}(\bm{n})$ and then summing the product
over $\bm{n}$.

Let us now consider the functional $\mathcal{M}'^{[0,t')}_{\bm{n}_t} 
\delta_{\bm{n},\bm{n}_t} \mathcal{M}_{\bm{n}_0}^{[0,t)}$.  
In analogy to the previous case, we easily arrive to
\begin{eqnarray}\fl
\label{PROOF5}
\E  \left(
\prod_{r=1}^{R}\mathcal{M}_{r-1}
\delta_{ \bm{n} , \bm{n}_R} \prod_{r'=1}^{R'}\mathcal{M}_{r'-1}^{'} 
\right)=
\E  \left(
\prod_{r=1}^{R}\media{\mathcal{M}_{r-1}}_{w}
\prod_{r'=1}^{R'-1}\media{\mathcal{M}_{r'-1}^{'}}_{w}
\media{\mathcal{M}_{R'-1}^{'}\delta_{ \bm{n} , \bm{n}_R}}_{w} 
\right),
\end{eqnarray}
where, recalling that 
$\underline{\bm{n}'_{t'_0}}=\underline{\bm{n}_{t_{R}}}$,
the weighted ``averages'' are given in terms of probabilities
$\underline{p_r}$ defined recursively as in Eq.~(\ref{p}) 
with $r=1,\ldots,R+R'-1$.
Let $P_{R+R'}$ and $\widetilde{P}_{R+R'}$ be the obvious extensions
of the distributions $P_R$ and $\widetilde{P}_{R}$ previously considered.
As before, by using Eqs.~(\ref{Pstorto}) and (\ref{p1}), 
for $r=1,\ldots,R+R'-1$,
we can realize the transformation 
$P_{R+R'} \rightarrow \widetilde{P}_{R+R'}$ recursively:
along the interval $[0,t)$ we sample 
$\widetilde{P}_{1},\widetilde{P}_{2},\ldots,\widetilde{P}_{R}$,
whereas along $[0,t')$ we sample 
$\widetilde{P}_{R+1},\widetilde{P}_{R+2},\ldots,\widetilde{P}_{R+R'}$,
obtaining the configurations 
$\underline{\widetilde{\bm{n}}_{0}},\underline{\widetilde{\bm{n}}_{1}}
,\ldots,
\underline{\widetilde{\bm{n}}_{R}},
\underline{\widetilde{\bm{n}}'_{0}},
\underline{\widetilde{\bm{n}}'_{1}}, \ldots,
\underline{\widetilde{\bm{n}}'_{R'-1}},
\mathcal{D} \underline{\widetilde{\bm{n}}'_{R'-1}}$.
Therefore, Eq.~(\ref{PROOF5}) transforms into
\begin{eqnarray}\fl
\label{PROOF6}
\E  \left(
\prod_{r=1}^{R}\mathcal{M}_{r-1}
\delta_{ \bm{n} , \bm{n}_R} \prod_{r'=1}^{R'}\mathcal{M}_{r'-1}^{'} 
\right) &=& 
\widetilde{\E}  \left(\prod_{r=1}^{R}
\media{\mathcal{M}_{\widetilde{\bm{n}}_{r-1}}^{[t_{r-1},t_r)}} 
\prod_{r'=1}^{R'-1}
\media{\mathcal{M}_{\widetilde{\bm{n}}'_{r'-1}}'^{[t_{r'-1},t_{r'})}} 
\right. \nonumber \\ && \left. \times \frac{1}{M}
\sum_{l=1}^M
\mathcal{M}_{\widetilde{\bm{n}}'^{(l)}_{R'-1}}'^{[t_{R'-1},t_{R'})(l)} 
\delta_{ \bm{n} , 
(\mathcal{R}^{R'}\mathcal{D} \underline{\widetilde{\bm{n}}_{R-1}})^{(l)} }
\right),
\end{eqnarray}
which yields Eq.~(\ref{RCORRD}) after multiplying 
$\mathcal{M}_{\bm{n}_0}^{[0,t)}\delta_{\bm{n},\bm{n}_t}$ by 
$O (\bm{n})$ and then summing over $\bm{n}$.
Note that in the r.h.s. of Eq.~(\ref{PROOF6}) there appears 
$\mathcal{R}^{R'}\mathcal{D} \underline{\widetilde{\bm{n}}_{R-1}}$
and not $\mathcal{D}\underline{\widetilde{\bm{n}}_{R-1}}$. 
Indeed, according to Eq.~(\ref{PROOF2}), 
in the last weighted average
$\media{\mathcal{M}_{R'-1}'\delta_{ \bm{n} , \bm{n}_R}}_{w}=
\sum_{l=1}^M 
\mathcal{M}_{R'-1}'^{(l)}\delta_{\bm{n}_R^{(l)},
\bm{n}}p'^{(l)}_{R+R'-1}$
there appear the probabilities $\underline{p'_{R+R'-1}}$
associated to the last time interval.

In general, in the reconfiguration procedure a weighted ``average''
\begin{equation}
\media{\mathcal{M}^{[t_{R-1},t_R)}_{{\bm{n}}_{R-1}}
f(\bm{n}_0,\bm{n}_1,\dots,\bm{n}_{R-1},\bm{n}_{R})}_w
\end{equation} 
will be substituted by the uniform ``average''
\begin{eqnarray}\fl
\label{fW}
\frac{1}{M}
\sum_{l=1}^M
\mathcal{M}^{[t_{R-1},t_R)(l)}_{\widetilde{\bm{n}}_{R-1}^{(l)}}
f((\mathcal{R}^{R-1}\underline{\widetilde{\bm{n}}_0})^{(l)},
(\mathcal{R}^{R-2}\underline{\widetilde{\bm{n}}_1})^{(l)},\dots,
(\mathcal{R}\underline{\widetilde{\bm{n}}_{R-1}})^{(l)},
(\mathcal{D}\underline{\widetilde{\bm{n}}_{R-1}})^{(l)}).
\end{eqnarray}
Equation (\ref{RCORROFF}) can be obtained in the same way as 
Eq.~(\ref{RCORRD}).
Finally, in the case of real times, Eq.~(\ref{RECONFIGURATIONi})
is immediately obtained by using for the local weights 
the quantities $| \mathcal{M}^{[it_{r-1},it_r)}_{\bm{n}_{t_{r-1}}} |$
and for the function $f(\cdot)$ of Eq.~(\ref{fW})
the product of the phase factors
\begin{eqnarray}
\label{phi}
f=
\prod_{r=1}^{R-1}
e^{ i \Phi^{[t_{r-1},t_r)}_{\bm{n}_{t_{r-1}}} }.
\end{eqnarray}

\section{Numerical results}
\label{Numerical results}
In this Section, we present some numerical applications of the algorithm
described above.
In principle, the reconfiguration scheme can be applied for any 
positive integer $R$.
However, we have observed optimal reconfiguration for 
$R\simeq\media{A}\rho t$, where $\media{A}$ is the average number of 
active links. 
This is what one expects as, in this case, the reconfiguration 
is repeated in the average at each jump, \textit{i.e.} 
as frequently as the stochastic dynamics dictates 
(see also Ref.~\cite{CAFFAREL}). 
In the simulations reported below, therefore, we always work 
with this approximately optimal number of reconfigurations.

The count of the active links and of the potential of a given 
configuration, quantities to be determined at each jump, 
is a core point of the algorithm.
Starting form a first count based on a systematic inspection of the
initial lattice configuration, 
we have implemented a local updating of these quantities.
In fact, when a jump occurs, the new Hubbard potential and the new 
number of active links are determined by the change 
of the lone occupations and of the lone links involved in the jump.  
The computational cost of a local update, which takes into account only
these relevant sites and links, is independent of the lattice size.
Also the reconfiguration procedure has been optimized by defining a 
non-negative integer, 
the replication multiplicity $\mu_r^{(i)}$, where $(i)$ is the
trajectory index and $\sum_{i=1}^M \mu_r^{(i)} = M$.
Configurations for which $\mu_r^{(i)}=0$ are substituted by 
those with $\mu_r^{(i)} >1$, whereas no operation is performed
for the trajectories with $\mu_r^{(i)}=1$, which are the largest
fraction of the whole set of $M$ trajectories.  
The efficiency of the resulting code can be figured out by the 
following example.
With an ordinary personal computer and without using importance sampling, 
we are able to simulate lattices with $40 \times 40$ sites with 800 
hard-core bosons  
obtaining the ground-state energy up to a relative error of the order of
$1 \%$ with 290 minutes of cpu time.
A detailed comparison of the efficiency of our EPRMC code with those 
implementing other Monte Carlo methods is beyond the
purposes of present work.
In the Appendix, we discuss the relative efficiency between 
EPRMC and GFMC or GFMCP.

\begin{figure}
\centering
\psfrag{x}[t][]{$t$}
\psfrag{y}[b][]{$\E (\mathcal{M}_{\bm{n}_0}^{[0,t)}) $} 
\includegraphics[width=0.75\columnwidth,clip]{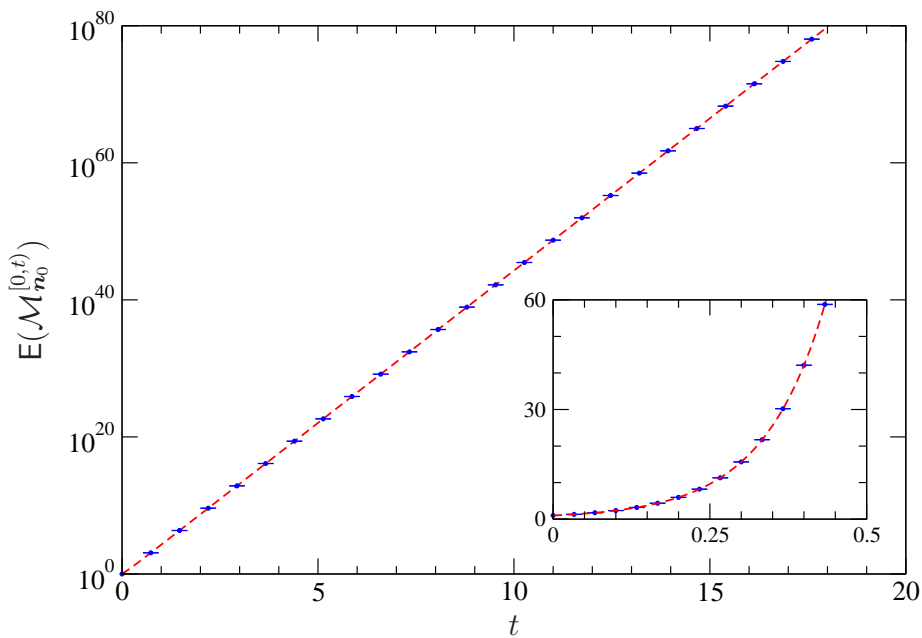}
\caption{
Expectation of the functional $\mathcal{M}_{\bm{n}_0}^{[0,t)}$ 
\textit{versus} the imaginary time $t$ for a hard-core boson Hubbard system 
in a $2 \times 3$ lattice at half filling 
with $\eta=1$, $\gamma=4$, and periodic boundary conditions.
The initial configuration is $\bm{n}_0= (1,1,1,0,0,0,1,1,0,1,0,0)$.
The Monte Carlo simulation (dots with error bars) was done with 
$M=2^{14}$ trajectories, $N=2^7$ samples, and $R=300$ reconfigurations.
Error bars correspond to one standard deviation evaluated from the $N$ samples.
The dashed line is the exact result from numerical diagonalization 
of the corresponding Hamiltonian.
In the inset we show the small time behavior.}  
\label{mc.fig1.eps}
\end{figure}

\begin{figure}
\centering
\psfrag{x}[t][]{$t$}
\psfrag{y}[b][]{$\mathrm{Re} \E (\mathcal{M}_{\bm{n}_0}^{[0,it)}) $,
$\mathrm{Im} \E (\mathcal{M}_{\bm{n}_0}^{[0,it)}) $ } 
\psfrag{Re}[][]{$\mathrm{Re}$}
\psfrag{Im}[][]{$\mathrm{Im}$}
\includegraphics[width=0.75\columnwidth,clip]{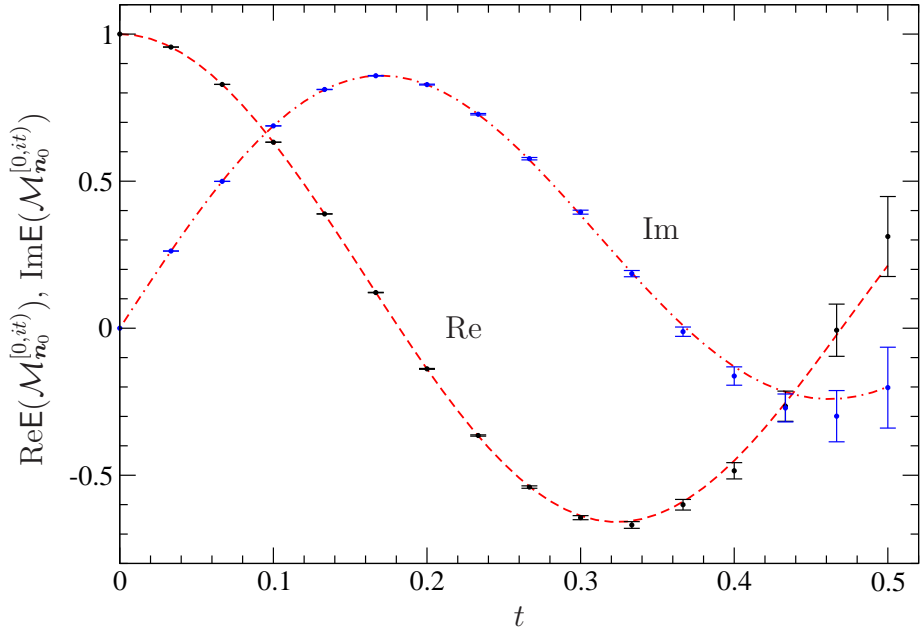}
\caption{
Expectation of the real and imaginary parts of the functional 
$\mathcal{M}_{\bm{n}_0}^{[0,it)}$ \textit{versus} the real time $t$ 
for the same system of Fig.~\ref{mc.fig1.eps}. 
The Monte Carlo simulation (dots with error bars) was done with 
$M=2^{20}$ trajectories, $N=2^7$ samples, and $R=15$ reconfigurations. 
Error bars correspond to one standard deviation evaluated from the $N$ samples.
The dashed (real part) and dot-dashed (imaginary part) lines are 
the exact results from numerical diagonalization.}
\label{mc.fig2.eps}
\end{figure}

\begin{figure}
\centering
\psfrag{x}[t][]{$t$}
\psfrag{y}[b][]{$\E (\mathcal{M}_{\bm{n}_0}^{[0,t)} \mathcal{H}(\bm{n}_t)) / 
\E (\mathcal{M}_{\bm{n}_0}^{[0,t)})$}
\includegraphics[width=0.75\columnwidth,clip]{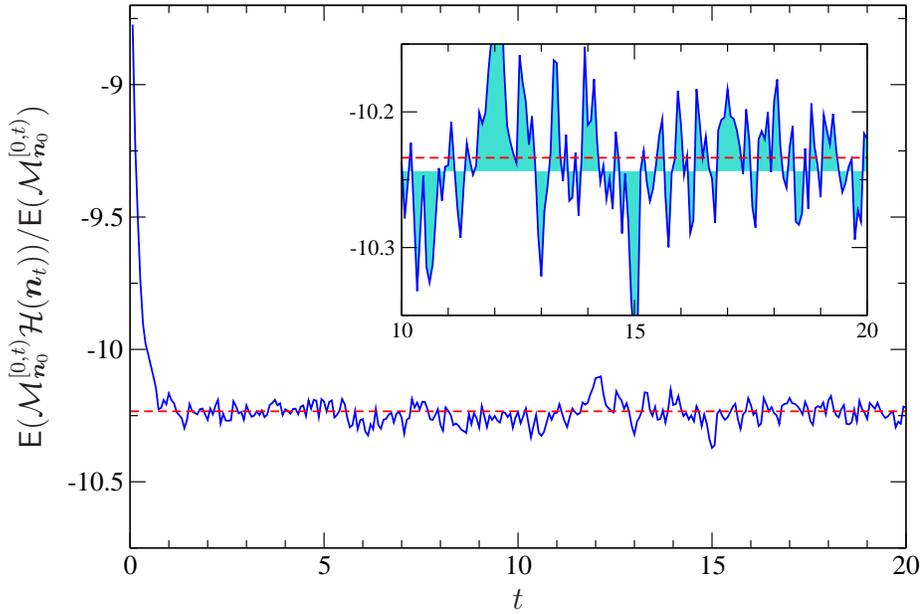}
\caption{
Local energy 
$\E (\mathcal{M}_{\bm{n}_0}^{[0,t)}\mathcal{H}(\bm{n}_t))/ 
\E (\mathcal{M}_{\bm{n}_0}^{[0,t)})$ 
\textit{versus} the imaginary time $t$ 
for the same system of Fig.~\ref{mc.fig1.eps}. 
The Monte Carlo simulation (solid line) was done with 
$M=2^{14}$, $N=1$, and $R=300$.
The straight dashed line is the exact energy $E_0=-10.233803$
obtained by diagonalization.
In the inset we evidence the difference between $E_0$ and the time average of 
$\E (\mathcal{M}_{\bm{n}_0}^{[0,t)}\mathcal{H}(\bm{n}_t)) / 
\E (\mathcal{M}_{\bm{n}_0}^{[0,t)})$ computed over the interval 
$10 \leq t \leq 20$ (opaque region baseline).}
\label{mc.fig3.eps}
\end{figure}

\begin{figure}
\centering
\psfrag{x}[t][]{$M$}
\psfrag{y}[b][]{$ [ \E (
\mathcal{M}_{\bm{n}_0}^{[0,t)}\mathcal{H}(\bm{n}_t)) / 
\E (\mathcal{M}_{\bm{n}_0}^{[0,t)}) ] / E_0 - 1$}
\includegraphics[width=0.75\columnwidth,clip]{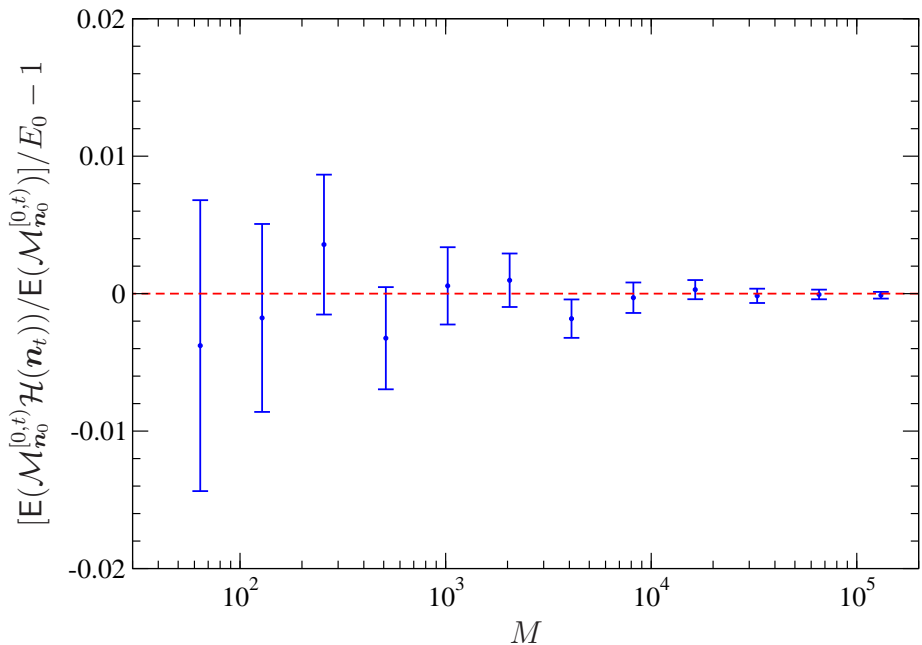}
\caption{
Relative error between the local energy 
$\E (\mathcal{M}_{\bm{n}_0}^{[0,t)}\mathcal{H}(\bm{n}_t)) / 
\E (\mathcal{M}_{\bm{n}_0}^{[0,t)})$
and the exact energy $E_0$ \textit{versus} the number $M$ of 
reconfigured trajectories for the same system of Fig.~\ref{mc.fig1.eps} 
with $N=2^7$, $t=5$, and $R=75$.
Error bars correspond to two standard deviations evaluated from the $N$ samples.
}
\label{mc.fig4.eps}
\end{figure}

\begin{figure}
\centering
\psfrag{x}[t][]{$\gamma$}
\psfrag{y}[b][]{$S(\pi,\pi)$}
\includegraphics[width=0.75\columnwidth,clip]{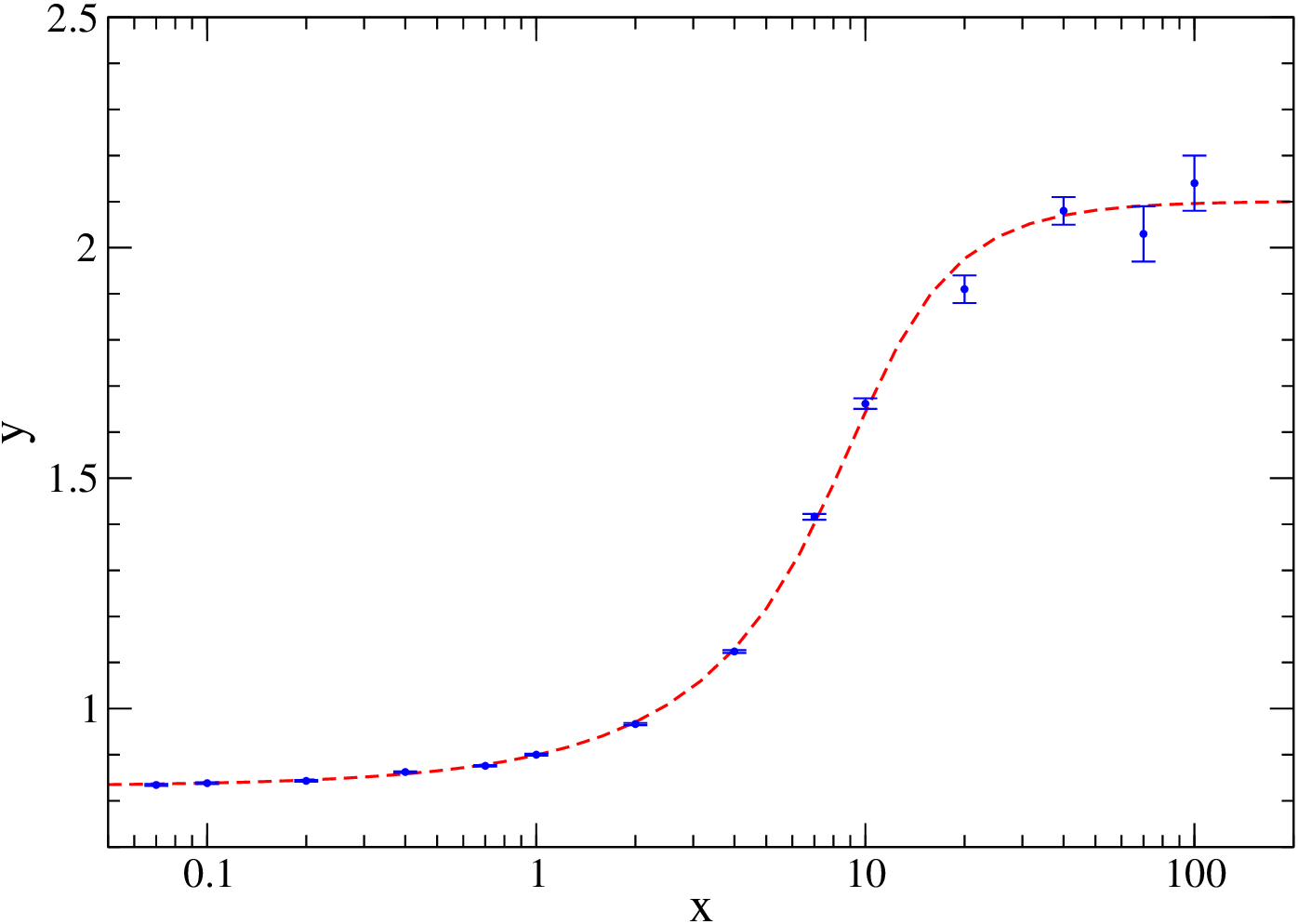}
\caption{
Spin-spin structure factor $S(q_x,q_y)$ at $q_x=q_y=\pi$ 
\textit{versus} the interaction strength $\gamma$ 
for the same system of Fig.~\ref{mc.fig1.eps}.
The dashed line is the exact result from numerical diagonalization of the
Hamiltonian whereas the dots with error bars are from a Monte Carlo simulation 
with $M=2^{14}$ ($M=2^{18}$ for $\gamma > 10$), $N=2^7$, $t=t'=3$, and $R=45$.
Error bars correspond to one standard deviation evaluated from the 
$N$ samples.
}
\label{mc.fig5.eps}
\end{figure}

In Figs. \ref{mc.fig1.eps}-\ref{mc.fig5.eps} we compare several quantities 
evaluated by the EPRMC algorithm with the corresponding exact results 
obtained by numerical diagonalization of the associated Hamiltonian. 
The system considered is a hard-core boson Hubbard model of small size,
namely a $2 \times 3$ lattice at half filling. 
The general purpose of these figures is to show the unbiased statistical 
convergence of the Monte Carlo data towards the exact values. 
No importance sampling is used in these first examples.
%%%%%%%%%%%%%%%%%%%%%%%%%%%%%%
%%%%%%%%%%%%%%%%%%%%%%%%%%%%%%
%%%%%%%%%%%%%%%%%%%%%%%%%%%%%%
%%%%%%%%%%%%%%%%%%%%%%%%%%%%%%
%%%%%%%%%%%%%%%%%%%%%%%%%%%%%%
%%%%%%%%%%%%%%%%%%%%%%%%%%%%%%

In Fig.~\ref{mc.fig1.eps} we show the expectation 
$\E (\mathcal{M}_{\bm{n}_0}^{[0,t)})$ as a function of the imaginary time $t$.
The agreement with the corresponding quantum matrix element
$\sum_{\bm{n}} \langle \bm{n}|e^{-Ht} | \bm{n}_0\rangle $ is excellent. 
The reconfiguration procedure is able to control completely the 
fluctuations growing with $t$ so that the error bars, 
negligible on the used scale, do not increase by increasing the time. 
Similar results are obtained for different initial configurations $\bm{n}_0$.

In Fig.~\ref{mc.fig2.eps} we show the expectation 
$\E (\mathcal{M}_{\bm{n}_0}^{[0,it)})$ as a function of the real time $t$.
Also in this case there is an exact statistical convergence towards
the quantum matrix element
$\sum_{\bm{n}} \langle \bm{n}|e^{-iHt} | \bm{n}_0\rangle $.
However, in this case the reconfiguration procedure is able to control only a 
part of the fluctuations, namely those related to the modulus
of the functional $\mathcal{M}_{\bm{n}_0}^{[0,it)}$.
The fluctuations associated to the corresponding phase factor 
make the convergence harder and harder for large times.

In Fig.~\ref{mc.fig3.eps} we show the behavior of the local energy
$\E (\mathcal{M}_{\bm{n}_0}^{[0,t)}\mathcal{H}(\bm{n}_t))/ 
\E (\mathcal{M}_{\bm{n}_0}^{[0,t)})$ as a function of the imaginary time $t$.
According to Eq.~(\ref{E02}), the local energy converges
to the ground-state energy of the system, $E_0$, for large times.
In fact, after an initial transient 
inversely proportional to the gap $E_1-E_0$,
the ratio $\E (\mathcal{M}_{\bm{n}_0}^{[0,t)}\mathcal{H}(\bm{n}_t))/ 
\E (\mathcal{M}_{\bm{n}_0}^{[0,t)})$, estimated with a finite number of 
trajectories $M$, fluctuates around an average value that is close but
different from $E_0$ (see inset of Fig.~\ref{mc.fig3.eps}).
However, if we increase $M$, as shown in Fig.~\ref{mc.fig4.eps}, 
the statistical accuracy increases and we obtain
an unbiased convergence toward $E_0$.

As an example of correlation functions, we studied 
the spin-spin structure factor
\begin{eqnarray}
\label{S}
S(q_x,q_y) =  \frac{1}{|\Lambda|}
\sum_{i,j \in \Lambda} 
e^{iq_x\left(x_i - x_j\right) + iq_y\left(y_i - y_j\right) }
\langle E_0 | S_i S_j  | E_0 \rangle,
\end{eqnarray}
where $S_i = c_{i\ua}^\dag c_{i\ua}^{} - c_{i\da}^{\dag} c_{i\da}^{}$ and
$x_i$ and $y_i$ are the coordinates of the $i$-th lattice point.
Note that the operators $S_iS_j$ are diagonal in the Fock space 
and can be evaluated by using Eq.~(\ref{RCORRD}).
In Fig.~\ref{mc.fig5.eps} we show $S(\pi,\pi)$ 
evaluated for different values of the interaction strength $\gamma$. 
In agreement with the exact results from numerical diagonalization,
$S(\pi,\pi)$ has a sharp transition between 
the $\gamma \to 0$ and $\gamma \to \infty$ asymptotic values.
This transition is expected to take place 
when the average kinetic and potential energies are of the same order,
\textit{i.e.},
for $\eta \media{A} \sim \gamma \media{N_{\ua\da}}$, where 
$\media{N_{\ua\da}}$ is the average number of doubly occupied sites.
For the system considered in Fig.~\ref{mc.fig5.eps}, we have
$\media{A}\simeq 15$ and $\media{N_{\ua\da}} \simeq 1.5$ so that the transition
is expected at $\gamma/\eta \sim 10$.
This is in agreement with the numerical results.

\begin{figure}
\centering
\psfrag{40x40}[][]{$40 \times 40$}
\psfrag{20x20}[][]{$20 \times 20$}
\psfrag{x}[t][]{$t$}
\psfrag{y}[b][]{$[ \E (\mathcal{M}_{\bm{n}_0}^{[0,t)} \mathcal{H}(\bm{n}_t)) 
/ \E (\mathcal{M}_{\bm{n}_0}^{[0,t)}) ] / |\Lambda| $}
\includegraphics[width=0.75\columnwidth,clip]{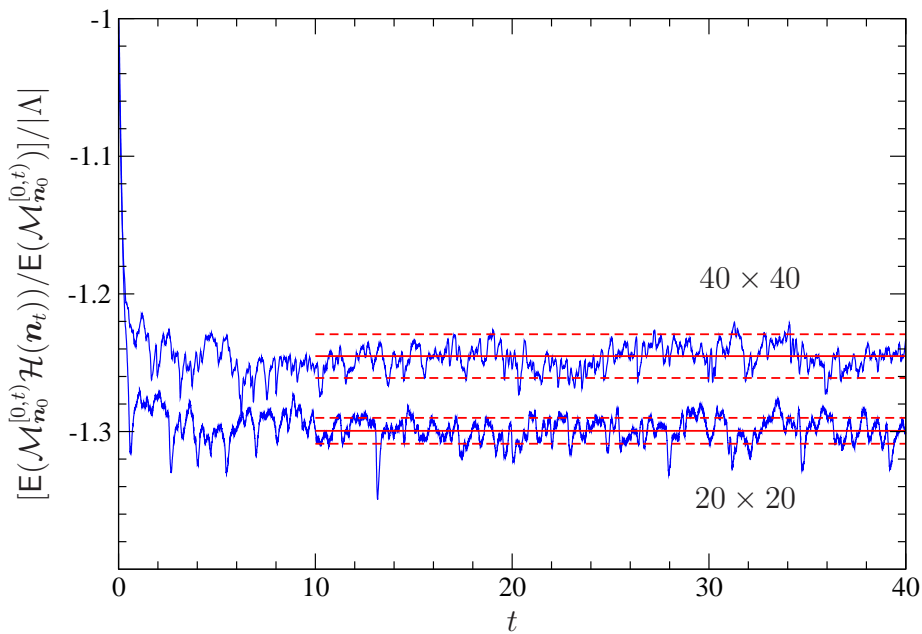}
\caption{Local energy per site
$[ \E (\mathcal{M}_{\bm{n}_0}^{[0,t)}\mathcal{H}(\bm{n}_t))/ 
\E (\mathcal{M}_{\bm{n}_0}^{[0,t)})] / |\Lambda|$ 
\textit{versus} the imaginary time $t$ 
for two different-size hard-core boson Hubbard systems
at half filling 
with $\eta=1$, $\gamma=4$, and periodic boundary conditions.
The Monte Carlo simulations (solid lines) were done with 
$M=2^{12}$, $R=2^{15}$, and $N=1$.
The straight solid lines are the time averages of the simulation results 
computed over the interval $10 \leq t \leq 40$ 
whereas the straight dashed lines indicate the relative standard deviations.
The simulations took 79 ($20 \times 20$ lattice) and 
290 ($40 \times 40$ lattice) minutes on a computer 
with a 2.40GHz Intel Xeon CPU.}
\label{mc.fig6.eps}
\end{figure}

In Fig.~\ref{mc.fig6.eps} 
we report simulations performed for
hard-core boson Hubbard systems of large size.
In particular, we show the local energy per site
$[ \E (\mathcal{M}_{\bm{n}_0}^{[0,t)}\mathcal{H}(\bm{n}_t))/ 
\E (\mathcal{M}_{\bm{n}_0}^{[0,t)})] / |\Lambda|$ 
as a function of the imaginary time $t$ for two lattices at half filling 
having size $20 \times 20$ and $40 \times 40$.
Note that the standard deviations of the fluctuations around the long-time
averaged value of 
$[ \E (\mathcal{M}_{\bm{n}_0}^{[0,t)}\mathcal{H}(\bm{n}_t))/ 
\E (\mathcal{M}_{\bm{n}_0}^{[0,t)})] / |\Lambda|$ 
provide an estimated relative error for $E_0/|\Lambda|$ of the order of $1 \%$.
This result is obtained with a moderate computational effort.
In Fig.~\ref{mc.fig6.eps} it is also evident an asymmetry of the 
fluctuations of the local energy around its mean value. 
This behavior is due to the reconfiguration procedure that ensures 
the invariance of the first statistical moment of $\mathcal{M}^{[0,t)}$ 
(or of related quantities) only.
%%%%%%%%%%%%%%%%%%%%%%%%%%%%%%
%%%%%%%%%%%%%%%%%%%%%%%%%%%%%%

We have performed simulations also for the Heisenberg model 
(\ref{HEISENBERG1}).
In this case, we used importance sampling with the following Jastrow-like
guiding state \cite{CALANDRASORELLA}
\begin{eqnarray}
\label{importance}
\langle\bm{n}|g\rangle\equiv
\exp \left[ {\frac{\alpha}{2}\sum_{i,j \in \Lambda}
\upsilon(\bm{r}_i-\bm{r}_j)\left(n_{i}-\frac{1}{2}\right)
\left(n_{j}-\frac{1}{2}\right)} \right],
\end{eqnarray} 
where $\bm{r}_i=(x_i,y_i)$, $\alpha$ is a real positive parameter, 
and the long range potential $\upsilon$ is defined as
\begin{eqnarray}
\upsilon(\bm{r})=\frac{2}{|\Lambda|}
\sum_{(q_x, q_y)\neq (0,0)}
e^{iq_x x + iq_y y }
\left[1-\sqrt{\frac{1+(\cos q_x+\cos q_y)/2}{1-(\cos q_x+\cos q_y)/2}}
~\right],
\end{eqnarray} 
the sum over $q_x$ and $q_y$ being extended over the
Brillouin zone ${2\pi/L,4\pi/L,\dots,2\pi}$, with $0$ excluded. 
From Eq.~(\ref{importance}) we have
\begin{eqnarray}
\label{ggg}
\frac
{\langle \bm{n}\oplus \bm{1}_{k} \oplus \bm{1}_{l}| g \rangle}
{\langle \bm{n}| g \rangle} &=&
\exp \Bigg[ \alpha\sum_{i \in \Lambda, ~i \neq k,~l}
\left(n_{i}-\frac{1}{2}\right)
[\left(n_{k}\oplus 1 - n_{k} \right)
\upsilon(\bm{r}_i-\bm{r}_k) 
\nonumber \\ &&
+\left(n_{l}\oplus 1 - n_{l} \right)
\upsilon(\bm{r}_i-\bm{r}_l)] \Bigg].
\end{eqnarray}
We assumed $\alpha=1.2$ as suggested in Ref.~\cite{CALANDRASORELLA}. 

\begin{figure}
\centering
\psfrag{x}[t][]{$t$}
\psfrag{y}[b][]{$[\E (\mathcal{M}_{g,\bm{n}_0}^{[0,t)} 
\mathcal{H}_g(\bm{n}_t)) / 
\E (\mathcal{M}_{g,\bm{n}_0}^{[0,t)})] / \left| \Lambda \right|$}
\includegraphics[width=0.75\columnwidth,clip]{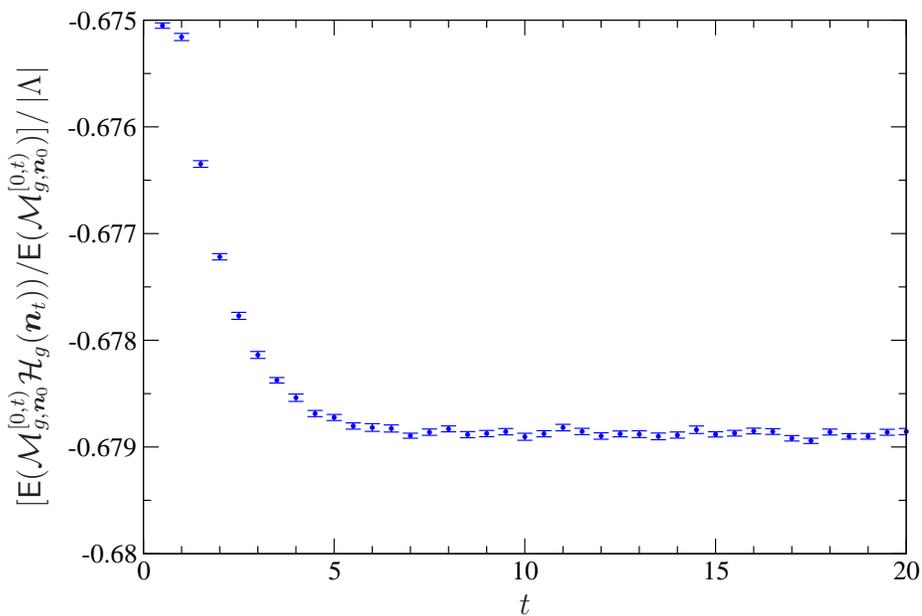}
\caption{Local energy per site
$[\E (\mathcal{M}_{g,\bm{n}_0}^{[0,t)}\mathcal{H}_g(\bm{n}_t))/ 
\E (\mathcal{M}_{g,\bm{n}_0}^{[0,t)})] / \left| \Lambda \right|$ 
\textit{versus} the imaginary time $t$ for a $6 \times 6$
Heisenberg system  with $\sum_{i=1}^{\left| \Lambda \right|} S_i^{z}=0$
and $J=1$.
The Monte Carlo simulation (dots with error bars) was done 
by using importance sampling with the guiding function (\ref{ggg}) 
and statistical parameters $M=2^{16}$, $N=2^6$, and $R=20$.
}
\label{mc.fig7.eps}
\end{figure}

In Fig.~\ref{mc.fig7.eps} we show the local energy per site
$[\E (\mathcal{M}_{g,\bm{n}_0}^{[0,t)}\mathcal{H}_g(\bm{n}_t))/ 
\E (\mathcal{M}_{g,\bm{n}_0}^{[0,t)})] / \left| \Lambda \right|$ 
as a function of the imaginary time $t$ for a $6 \times 6$
Heisenberg system having $\sum_{i=1}^{\left| \Lambda \right|} S_i^{z}=0$.
The amplitude of the error bars shown in Fig.~\ref{mc.fig7.eps}
is considerably reduced with respect to the value that one would obtain
without using importance sampling. 
We also stress that the dynamics shown in \ref{mc.fig7.eps} is
relative to the Hamiltonian $H_g$ modified by the chosen guiding function
$|g\rangle$.
%%%%%%%%%%%%%%%%%%%%%%%%%%%%%%
%%%%%%%%%%%%%%%%%%%%%%%%%%%%%%
%%%%%%%%%%%%%%%%%%%%%%%%%%%%%%
\begin{figure}
\centering
\psfrag{x}[t][]{$t'$}
\psfrag{y}[b][]{$\sqrt{ 3
[\E (\mathcal{M}'^{[0,t')}_{g,\bm{n}_t} S^z_{\pi\pi} (\bm{n}_t) 
\mathcal{M}_{g,\bm{n}_0}^{[0,t)} ) /
\E (\mathcal{M}_{g,\bm{n}_0}^{[0,t+t')})] /|\Lambda| }$}
\psfrag{4x4}[][]{$4 \times 4$}
\psfrag{6x6}[][]{$6 \times 6$}
\psfrag{8x8}[][]{$8 \times 8$}
\includegraphics[width=0.75\columnwidth,clip]{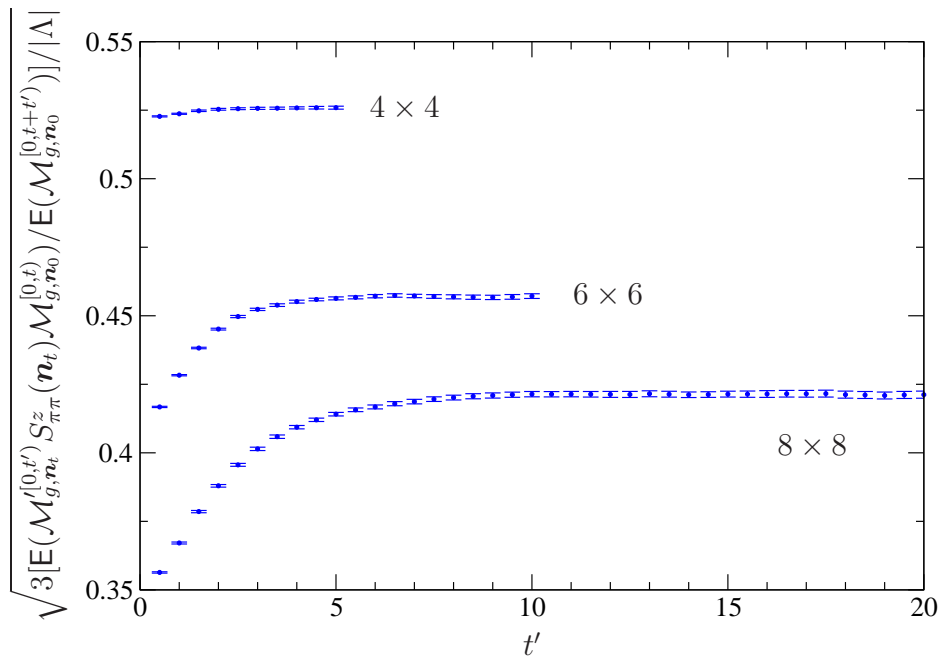}
\caption{
Local staggered magnetization 
$\{ 3
[\E (\mathcal{M}'^{[0,t')}_{g,\bm{n}_t} S^z_{\pi\pi} (\bm{n}_t) 
\mathcal{M}_{g,\bm{n}_0}^{[0,t)} ) / 
\E (\mathcal{M}_{g,\bm{n}_0}^{[0,t+t')})] \linebreak /|\Lambda| \}^{1/2}$,
defined in terms of the operator (\ref{Szpp}), 
\textit{versus} the imaginary time $t'$ for 
different-size Heisenberg systems 
with $\sum_{i=1}^{\left| \Lambda \right|} S_i^{z}=0$ and $J=1$.
The Monte Carlo simulations (dots with error bars) were done 
by using importance sampling with the guiding function (\ref{ggg}) 
with $\alpha=1.2$ and parameters   
$t=5$, $M=2^{16}$, $N=22$, and $R=10$ in the $4 \times 4$ case,
$t=10$, $M=2^{16}$, $N=22$, and $R=20$ in the $6 \times 6$ case,
$t=20$, $M=2^{17}$, $N=22$, and $R=40$ in the $8 \times 8$ case.
}
\label{mc.fig8.eps}
\end{figure}
%%%%%%%%%%%%%%%%%%%%%%%%%%%%%%

In Fig.~\ref{mc.fig8.eps} we show the 
local staggered magnetization
$\{ 3
[\E (\mathcal{M}'^{[0,t')}_{g,\bm{n}_t} S^z_{\pi\pi} (\bm{n}_t) 
\mathcal{M}_{g,\bm{n}_0}^{[0,t)} ) \linebreak / 
\E (\mathcal{M}_{g,\bm{n}_0}^{[0,t+t')})] /|\Lambda| \}^{1/2}$
evaluated in Heisenberg systems of different size
as a function of the imaginary time $t'$ and 
for a large value of the other imaginary time $t$. 
Here $S^z_{\pi\pi} (\bm{n}) = \langle \bm{n} | S^z_{\pi\pi} |\bm{n} \rangle$ 
is the quantum expectation in the Fock state $\bm{n}$ of the diagonal
operator
\begin{eqnarray}
\label{Szpp}
S^z_{\pi\pi} =  \frac{1}{|\Lambda|}
\sum_{i,j \in \Lambda} 
e^{i \pi \left(x_i - x_j\right) + i \pi \left(y_i - y_j\right) }
S_i^z S_j^z.
\end{eqnarray}
As noticed in the case of Fig.~\ref{mc.fig7.eps}, also
the dynamics of the local staggered magnetization shown in 
Fig.~\ref{mc.fig8.eps} is relative to the Hamiltonian $H_g$ 
modified by the guiding function (\ref{ggg}).
The asymptotic values of the local staggered magnetization
reached for $t'$ large
are in agreement with those obtained with different Monte Carlo
algorithms \cite{CALANDRASORELLA,SANDVIK}.
The statistical errors shown in Fig.~\ref{mc.fig8.eps} 
can be reduced by a factor about $10$ 
by exploiting the covariance between the local estimators
for $S^z_{\pi,\pi}$ and $E_0$, as explained in \cite{SANDVIK}.

\begin{figure}
\centering
\psfrag{x}[t][]{$t'$}
\psfrag{y}[b][]{$\sqrt{ 3
[\E (\mathcal{M}'^{[0,t')}_{g,\bm{n}_t} S^z_{\pi\pi} (\bm{n}_t) 
\mathcal{M}_{g,\bm{n}_0}^{[0,t)} ) /
\E (\mathcal{M}_{g,\bm{n}_0}^{[0,t+t')})] /|\Lambda| }$}
\psfrag{a=0.6}[][]{$\alpha=0.6$}
\psfrag{a=1.2}[][]{$\alpha=1.2$}
\psfrag{a=1.8}[][]{$\alpha=1.8$}
\includegraphics[width=0.75\columnwidth,clip]{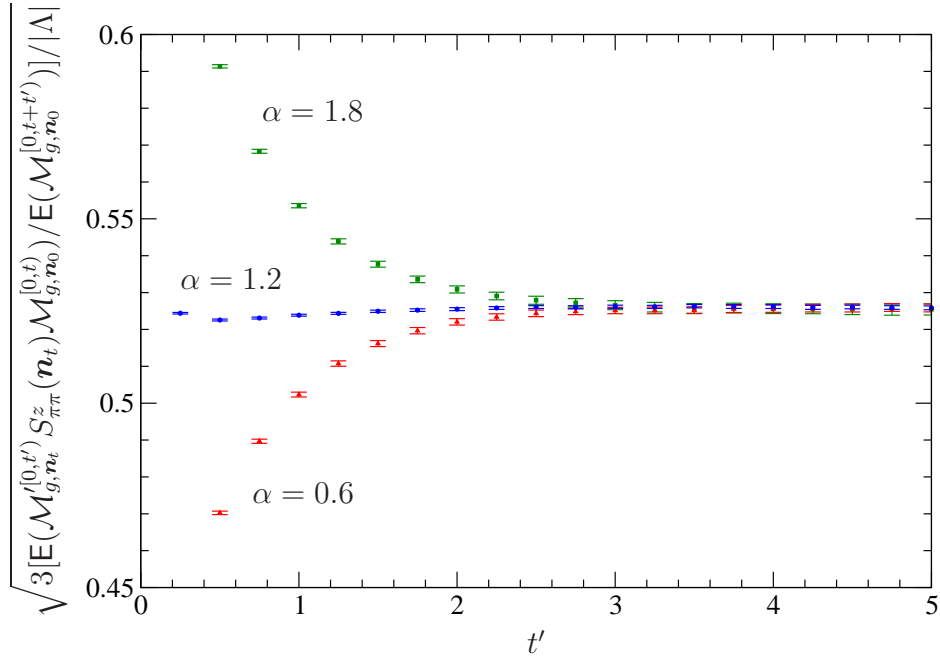}
\caption{
Local staggered magnetization \textit{versus} the imaginary time $t'$ 
for the $4 \times 4$ Heisenberg system of Fig.~\ref{mc.fig8.eps}. 
The three curves were obtained with different values of the parameter 
$\alpha$ defining the importance sampling function (\ref{ggg}).
The other parameters were
$t=5$, $M=2^{16}$, $N=22$, and $R=20$.
}
\label{mc.fig9.eps}
\end{figure}

Finally, in Fig.~\ref{mc.fig9.eps} we provide an example of how the 
local expectation values shown in Fig.~\ref{mc.fig8.eps}
depend on the parameter $\alpha$ of the guiding function (\ref{ggg}).
For different values of $\alpha$ the local expectations have different
evolutions determined by $H_{g(\alpha)}$, however, 
as stated in Section \ref{Importance sampling} for a general guiding function, 
they all converge to the quantum expectation of $S^z_{\pi\pi}$
in the ground state of $H$. 
In agreement with Ref.~\cite{CALANDRASORELLA}, the value 
$\alpha=1.2$ is close to the optimal choice, 
which provides smallest fluctuations
and minimal evolution with respect to the asymptotic values. 
%%%%%%%%%%%%%%%%%%%%%%%%%%%%%%
%%%%%%%%%%%%%%%%%%%%%%%%%%%%%%

In Figs. \ref{mc.fig1.eps} and \ref{mc.fig2.eps} we have shown that 
the imaginary- and real-time evolution of the expectation of the 
basic functional $\mathcal{M}_{\bm{n}_0}^{[0,t)}$ coincides
with that of the corresponding quantum matrix element
$\sum_{\bm{n}} \langle \bm{n}|e^{-Ht} | \bm{n}_0\rangle $.
Of course, a similar behavior is general.
Even if not shown explicitly, in all the considered examples
the evolution of the relevant time-dependent probabilistic expectations 
coincides with that of the corresponding time-dependent quantum 
correlations functions.

\section{Conclusions}  
\label{Conclusions}  
We have exploited an exact probabilistic representation of the
quantum dynamics in a lattice to derive a Monte Carlo algorithm,
named EPRMC, for which standard fluctuation control techniques
like reconfigurations and importance sampling have been adapted 
and rigorously proved.
This exact representation holds for both imaginary and real times,
even if in the latter case only a partial fluctuation control is possible
so that reliable statistical simulations are limited to short times. 

Monte Carlo algorithms, like GFMC or GFMCP, provide similar representations 
of the evolution operator, which are affected, however, 
by a systematic error $\varepsilon$ 
controlled by the number of iterations performed.
With respect to these approximated methods,
EPRMC gives an efficiency gain proportional to the accuracy 
$\varepsilon^{-1}$.

\section*{Acknowledgments}
This work was supported in part by
Cofinanziamento MIUR protocollo 2002027798\_001.

\section*{Appendix}
In this Appendix, we calculate
the relative efficiencies of GFMC and EPRMC methods.
Both the methods have the 
aim to sample the operator $e^{-Ht}$ for $t$ large. 
We can write
\begin{eqnarray}
U(t)= e^{-Ht} \sim e^{-E_0t}, \qquad \mbox{for } t \gg \overline{t},
\end{eqnarray}
where $\overline{t}$ is the characteristic time to filter out
the excited states $E_1,E_2,\ldots$ with respect to the ground
state $E_0$,
\begin{eqnarray}
\overline{t}=\frac{1}{E_1-E_0}.
\end{eqnarray}
As explained in the introduction, GFMC samples the operator 
$(1-Ht/N)^N$ whereas EPRMC samples directly the operator $e^{-Ht}$.
Since $\lim_{N\rightarrow \infty} (1+x/N)^N=e^{x}$,
GFMC $\rightarrow$ EPRMC as the number
of iterations $N$ in the GFMC method grows.
However, for a finite value of $N$, GFMC remains affected by a 
systematic error.
We are interested in evaluating the critical
value of $N$ above which this error becomes smaller than a given value.
Let us consider the relative difference
\begin{eqnarray}
\label{eff}
f_N(x)=\frac{e^x-(1+\frac{x}{N})^N}{e^x}.
\end{eqnarray}
By using 
\begin{eqnarray}
\log(1+y)=\sum_{k=1}^{\infty}\frac{(-1)^{k+1}}{k}y^k,
\end{eqnarray}
Eq.~(\ref{eff}) becomes
\begin{eqnarray}
\label{eff1}
f_N(x)=(1-e^{-\frac{x^2}{2N}+\frac{x^3}{3N^2}-\dots}).
\end{eqnarray}
For concreteness, let us put $x=-E_0t$ in Eq.~(\ref{eff1}).
If we require that the relative error is $f_N(-E_0t) = \varepsilon \ll 1$, 
then we must have $N \ge N_t(\varepsilon)$, where
\begin{eqnarray}
N_t(\varepsilon) = \frac{E_0^2t^2}{2 \varepsilon }.
\end{eqnarray}

In conclusion, $N_t(\varepsilon)$ is the number of steps needed in GFMC 
to sample the operator $e^{-Ht}$ for $t$ large 
with a relative error equal to $\varepsilon$.
On the other hand,
the number of steps needed in EPRMC to sample $e^{-Ht}$ for $t$ 
large is given by the average number of jumps that, 
when an optimal reconfiguration scheme is chosen as discussed
in Section \ref{Numerical results}, 
coincides with the number of reconfigurations $R_t$
\begin{eqnarray}
R_t=\media{A}\eta t\simeq E_0^{(0)}t,
\end{eqnarray}
where $\media{A}$ is the average number of active links 
and $E_0^{(0)}$ is the ground-state energy in the non-interacting case.
Therefore, the relative efficiency of EPRMC with respect to GFMC 
is given by the ratio
\begin{eqnarray}
\label{RATIO}
\frac{N_t(\varepsilon)}{R_t}= \frac{E_0^2t}{2E_0^{(0)} \varepsilon }.
\end{eqnarray}
We see that the superiority of EPRMC grows by
increasing the time $t$ or increasing the accuracy $\varepsilon^{-1}$ 
required in GFMC.
In particular for $t=\overline{t}$ we have 
\begin{eqnarray}\label{RATIO1}
\frac{N_{\overline{t}}(\varepsilon)}{R_{\overline{t}}}= 
\frac{E_0^2}{2E_0^{(0)}(E_1-E_0)  \varepsilon}.
\end{eqnarray}

It is clear that if, instead of GFMC, we consider the GFMCP method, 
the efficiency ratio (\ref{RATIO}) changes.
In fact, any step in GFMCP, on the average, amounts to $\media{n_s}$
elementary GFMC steps, where roughly $\media{n_s}=\media{A}\eta t$ 
\cite{PRESILLA}. Thus, in GFMCP the number of steps 
needed to sample the operator $e^{-Ht}$ for $t$ large 
with a relative error $\varepsilon$
is reduced to 
$N_{t}(\varepsilon)=E_0^2t/(2E_0^{(0)}\varepsilon)$ so that
the relative efficiency of EPRMC with respect to GFMCP 
is given by the ratio
\begin{eqnarray}
\label{RATIOP}
\frac{N_t(\varepsilon)}{R_t}= \left( \frac{E_0}{E_0^{(0)}} \right)^2
\frac{1}{2 \varepsilon }.
\end{eqnarray}
This ratio does not depend on $t$ anymore but remains proportional
to the accuracy required in GFMCP.


\section*{References}
\begin{thebibliography}{99}
\bibitem{LINDEN} von der Linden W 1992 \textit{Phys. Rep.} \textbf{220} 53
\bibitem{DAJLS} De Angelis G F, Jona-Lasinio G and Sirugue M 1983 \textit{J. Phys.} A {\bf 16} 2433
\bibitem{DJS} De Angelis G F, Jona-Lasinio G and Sidoravicius V 1998 \textit{J. Phys.} A {\bf 31} 289
\bibitem{PRESILLA} Beccaria M, Presilla C, De Angelis G F and Jona Lasinio G 1999 \textit{Europhys. Lett.} {\bf 48} 243
\bibitem{PROST} M. Ostilli and C. Presilla 2004 New J. Phys. \textbf{6} 107
\bibitem{CEPERLEYKALOS} Ceperley D M and Kalos M H 1992, in 
\textit{Monte Carlo Methods in Statistical Physics}, 
K. Binder Editor (Springer-Verlag, Heidelberg)
\bibitem{TRIVEDICEPERLEY} Trivedi N and Ceperley D M 1990 
\textit{Phys. Rev.} B \textbf{41} 4552
\bibitem{HETHERINGTON} Hetherington J H 1984 \textit{Phys. Rev. A} 
{\bf 30} 2713.
\bibitem{CALANDRASORELLA} Calandra Buonaura M and Sorella S 1998 \textit{Phys. Rev.} B \textbf{57} 11446
\bibitem{BECCARIA} Beccaria M 2000 \textit{Eur. Phys. J. C} {\bf 13} 357
\bibitem{HUBBARD} Hubbard J 1963 \textit{Proc. R. Soc. A} {\bf 276} 238
\bibitem{MATSUBARA} Matsubara T and Matsuda H 1956 \textit{Prog. Theor. Phys.} 
{\bf 16} 569
\bibitem{NOTE} Note that in Ref.~\cite{PRESILLA}
a more fundamental family of Poisson processes and 
stochastic functional are first introduced.
Equation (\ref{FORMULA CI})
is then derived as an effective algorithm.
\bibitem{PRESILLA1} Beccaria M, Presilla C, De Angelis G F and 
Jona Lasinio G 2001 \textit{Int. J. Mod. Phys.} {\bf 15} 1740
\bibitem{CAFFAREL} Assaraf R, Caffarel M and Khelif A 2000 
\textit{Phys. Rev. E} {\bf 61} 4566
\bibitem{SANDVIK} Sandvik A 1997 \textit{Phys. Rev.} B {\bf 56} 11678

\end{thebibliography}
\end{document}